\documentclass[aps,pre,preprint]{revtex4-2}
\usepackage{amsmath}
\usepackage{graphicx}
\usepackage{bbm}
\usepackage{subfigure}
\usepackage[caption=false]{subfig}
\begin{document}
\title{Diffusive Decay of Collective Quantum Excitations in Electron Gas}
\author{M. Akbari-Moghanjoughi}
\affiliation{Faculty of Sciences, Department of Physics, Azarbaijan Shahid Madani University, 51745-406 Tabriz, Iran\\ massoud2002@yahoo.com}

\begin{abstract}
In this work the multistream quasiparticle model of collective electron excitations is used to study the energy-density distribution of collective quantum excitations in an interacting electron gas with arbitrary degree of degeneracy. Generalized relations for the probability current and energy density distributions is obtained which reveals a new interesting quantum phenomenon of diffusive decay of pure quasiparticle states at microscopic level. The effects is studied for various cases of free quasiparticles, quasiparticle in an infinite square-well potential and half-space collective excitations. It is shown that plasmon excitations have the intrinsic tendency to decay into equilibrium state with uniform energy density spacial distribution. It is found that plasmon levels of quasipaticle in a square-well potential are unstable decaying into equilibrium state due to the fundamental property of collective excitations. The decay rates of pure plasmon states are determined analytically. Moreover, for damped quasiparticle excitations the non-vanishing probability current divergence leads to imaginary energy density resulting in damping instability of energy density dynamic. The pronounced energy density valley close to half-space boundary at low level excitations predicts attractive force close to the surface. Current research can have implications with applications in plasmonics and related fields. Current analysis can be readily generalized to include external potential and magnetic field effects. 
\end{abstract}

\date{\today}

\maketitle
\newpage

\section{Theoretical Background}

Collective excitations play critical role in natural phenomena. They produce large-scale patterns such as solitons, dispersive and shock waves and even chaotic behavior in most cases \cite{krall}. The plasma state definition strictly relies on the collective nature of oscillation of charged species \cite{chen}. Large class of instabilities arise due to delicate interplay of collective interactions among different species in plasmas via the dispersion and dissipation. However, as compared to other states of matter plasma occupies a large span of density-temperature regime making almost $99$ percent of the observable universe to consist of plasmas. In the two extreme density limits, on the other hand, the definition of plasmas (due to collective nature of act) becomes obscure. At very low concentration the number of particles inside the Debye sphere become insufficient for collective effects to manifest. In such a case only single particle aspects of phenomena become visible. At the other extreme, when the number density of charged species increases unboundedly for a given temperature, the plasmon blackout effect comes into play and prohibits collective effects. Therefore, collective quantum plasma, which arise due to definition when the de Broglie's wavelength compares to the inter-particle spacing, requires a more delicate definition in terms of plasma parameters. Such a characteristic definition requires a brand new equation of state (EoS) which has been discussed elsewhere \cite{akbnew1,akbnew2}.    

Quantum mechanics, by its virtue, has been originally designed to study the single-particle effects \cite{es,de}. However, its built-in statistical prediction ability leads to anomalous single-particle aspects which due to Copenhagen indeterministic-type interpretations has baffled scientists for many years and even led to rejection by pioneers in the field including Einstein with the famous quote: "God does not play dice". The apparent inconsistencies in physical interpretations of quantum theory had consequently led to appearance of deterministic theories such as de Broglie-Bohm pilot-wave and Madelung's fluid-like formalism. Due to inherent ability in statistical prediction beside simplicity, which has brought quantum mechanics enormous success in many fields of modern physics, the theory has been effectively used to incorporate many body effects by original works of Wigner and Moyal \cite{wigner,tat,moyal,groen}. The development of many body formulation in collective quantum theory of plasmas is indebted to many pioneering works \cite{fermi,bohm,bohm1,bohm2,pines,levine,klim1,5,6,7,8,9,50,51,52,53,lind}. There has been many successful recent developments in the fields of quantum kinetic \cite{mannew,zaman}, density functional \cite{dft} and quantum hydrodynamic theories \cite{fhaas} which have been explored through recent decade \cite{man2,shuk1,manfredi,haas1,brod1,mark1,man3,asenjo,brod2,mark2,stern,ydj1,ydj2,dub1,shuk2,shuk3,shuk4,mark3,sarma,brod3,mold1,mold2,sm,akbrevisit,akbnat1,akbnat2}. However, due to intrinsic complexity of many body quantum effects, detailed understanding of quantum effects has faced fundamental difficulties \cite{mark,haug,gardner}.     

Plasmonics is a newly emerging field of applied science relying on the understanding of collective quantum effects. Plasmonic research due to appealing technological applications has attracted growing interest in past few years. Plasmonic devices can find applications in optical emitters \cite{and}, plasmon focusing \cite{stock}, nanoscale waveguiding \cite{qui} and optical antennas \cite{muh}, nanoscale swiches \cite{kar} and plasmonic lasers \cite{oul}. Moreover, collective plasma excitations may become a more efficient way in solar energy conversion leading towards improved photovoltaic and catalytic designs by collective energy transport \cite{cesar,jac}. In current research we explore some new aspects of quasiparticle model of collective electron excitations concerning diffusion of pure collective electron states which can have important implications to the fields of nanoplasmonics. The phase space evolution of quasiparticle excitations in electron gas has been recently studied using this model \cite{akbnew}. We briefly discuss the model and the plasmon dispersion properties in Sec. II. Generalized probability current definition for damped collective excitations is given in Sec. III. The energy density of interacting electron gas is derived in Sec. IV. Application of the model to free quasiparticle, quasiparticle in a square-well potential and damped quasiparticle are respectively presented in Secs. V, VI and VII and conclusions are drawn in Sec. VIII.

\section{Quasiparticle Model and Dispersion Relation}

In this section we briefly introduce the quasiparticle model of collective excitations in electron gas using multistream model and review the energy band structure of damped plasmon excitations. Despite conventional many body theories in the multistream model electrons are localized in momentum space couples to each other via the Poisson equation. By discarding the probability concept of finding electrons, one avoids the redundant mean field approximations in order to solve the many body problem. We picture each electron as a quantum stream governed by universal Hamiltonian which encompasses all essential ingredients operating on individual particles in the system. The act of this hamiltonian is through the following single-particle Schr\"{o}dinger equation  
\begin{equation}\label{sch}
i\hbar \frac{{\partial {{\cal{N}}_j}({\bf{r}},t)}}{{\partial t}} = {\cal{H}}{{\cal{N}}_j}({\bf{r}},t),
\end{equation}
where $j$ denotes the stream number and the generalized Hamiltonian is ${\cal{H}} = - (\hbar^2/2m)\Delta  - (\kappa \cdot \nabla) - e\phi ({\bf{r}}) + \mu$ which, respectively from the left, includes the kinetic, damping, electrostatic potential and chemical potential operators. Here, $\kappa$ is parameter to introduce the damping effect in a phenomenological manner and contains the missing dimensional parameters in this term. Although, (\ref{sch}) is called the single-particle, the electrons are related through acting of the universal Hamiltonian operator which is itself under the influence of each particle via field equation such as Poisson's relation and appropriate equation of states (EoS) which relate given potentials in the Hamiltonian to the local density of electrons. Therefore, the wavefunction ${\cal{N}}_j$ describes not only the dynamics of given $j$-th electron in the system, but also its relation to others. The quantum electron streams are coupled via the Poisson's relation (assuming a known EoS) as follows     
\begin{equation}\label{poi}
\Delta \phi ({\bf{r}}) + 2(\kappa \cdot \nabla ) \phi ({\bf{r}})= 4\pi e\left[ {\sum\limits_{j = 1}^{N} {{{{\cal{N}}_j}({\bf{r}},t){\cal{N}}_j^*({\bf{r}},t)} - {n_0}} } \right],
\end{equation}
in which $n_0$ denotes the number density of the static neutralizing background charge. The local electron number density uses the standard definition in terms of wavefunction, $n = \sum\limits_{j = 1}^N {{{{\cal{N}}_j}({\bf{r}},t){\cal{N}}_j^*({\bf{r}},t)}}$. We now define the multistream wavefunction as ${\cal{N}}({\bf{r}},t) = \sum\limits_{j = 1}^N {{{\cal{N}}_j}({\bf{r}},t)}$ which is the analogous counterpart of many body wavefunction in standard treatments. However, in the multistream model the summation form replaces the products of single particle wavefunctions which causes main difficulties in many body treatments. Using this new definition one arrives at the following coupled system of differential equation for quasiparticle excitations  
\begin{subequations}\label{qp}
\begin{align}
&i\hbar\frac{{\partial {\cal {N}}({\bf{r}},t)}}{{\partial t}} =  - \frac{\hbar^2}{2m}\Delta {\cal {N}}({\bf{r}},t) - 2(\kappa  \cdot \nabla ){\cal {N}}({\bf{r}},t) - e\phi ({\bf{r}}){\cal {N}}({\bf{r}},t) + \mu {\cal {N}}({\bf{r}},t),\\
&\Delta \phi ({\bf{r}}) + 2(\kappa \cdot \nabla ) \phi ({\bf{r}}) = 4\pi e\left[ {{{\cal{N}}({\bf{r}},t){\cal{N}}^*({\bf{r}},t)} - \sum\limits_{k \ne j}^N {{{\cal{N}}_k}({\bf{r}},t){\cal{N}}_j^*({\bf{r}},t)} - {n_0}} \right].
\end{align}
\end{subequations}

In the limit of large electron number, ($N\gg 1$), the second term in rhs of (\ref{qp}) vanishes due to phase mixing effect and one arrive at the following simplified system \cite{akbnew1}
\begin{subequations}\label{eqp}
\begin{align}
&i\hbar\frac{{\partial {\cal {N}}({\bf{r}},t)}}{{\partial t}} =  - \frac{\hbar^2}{2m}\Delta {\cal {N}}({\bf{r}},t) - 2(\kappa  \cdot \nabla ){\cal {N}}({\bf{r}},t) - e\phi ({\bf{r}}){\cal {N}}({\bf{r}},t) + \mu {\cal {N}}({\bf{r}},t),\\
&\Delta \phi ({\bf{r}}) + 2(\kappa \cdot \nabla ) \phi ({\bf{r}}) = 4\pi e\left[ {{{\cal{N}}({\bf{r}},t){\cal{N}}^*({\bf{r}},t)} - {n_0}} \right].
\end{align}
\end{subequations}
Note that the summation in (\ref{qp}) which includes the mixed states can become important in few-body quantum systems and other electron interaction effects. In current analysis however we ignore the electron exchange and correlation effect which stand in the next order of importance.  

\begin{figure}[ptb]\label{Figure1}
\includegraphics[scale=0.6]{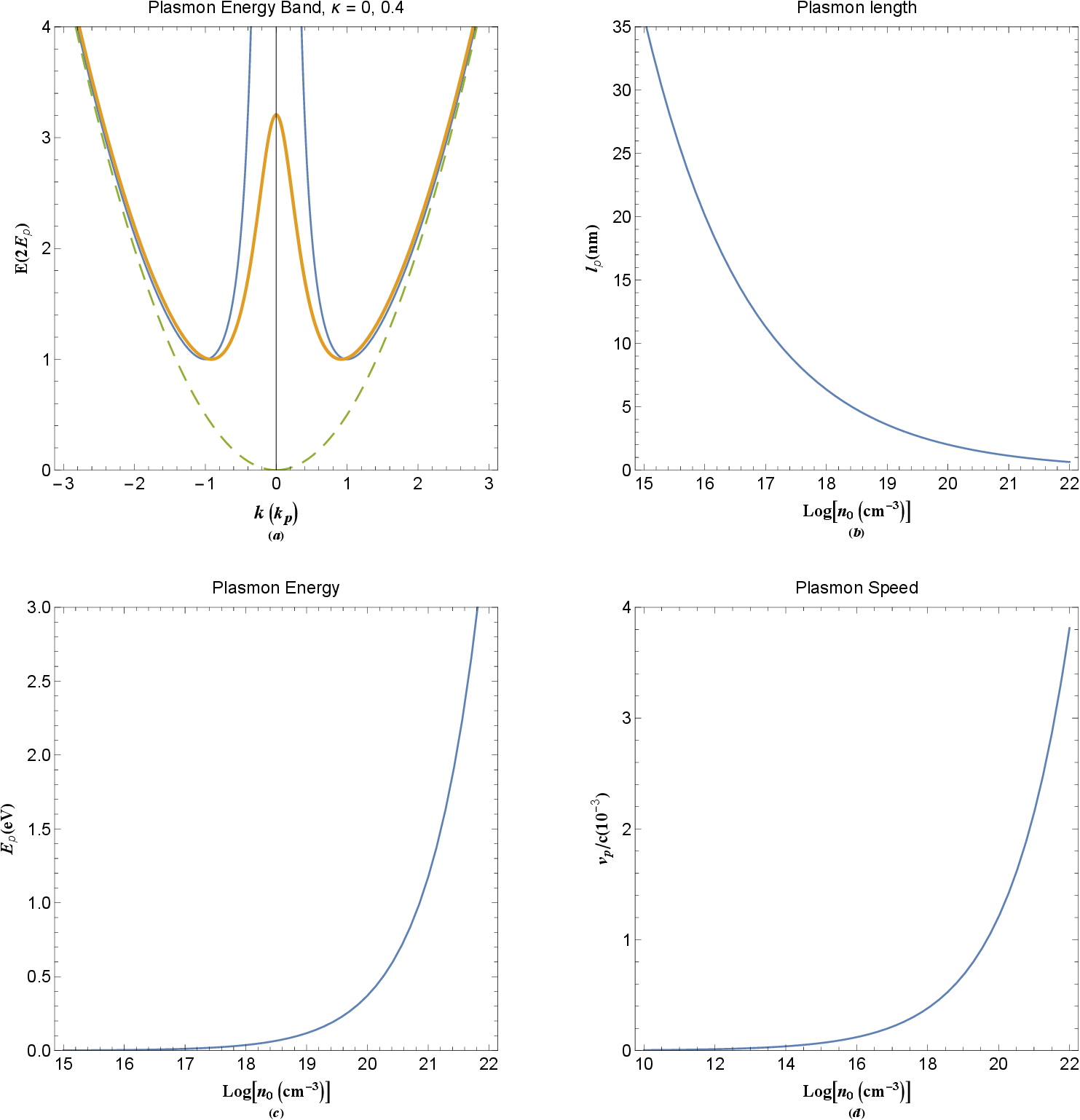}\caption{(a) Energy dispersion of undamped (thin solid curve), damped (thick curve) and free electron (dashed curve) excitations. (b) Variation of plasmon length versus free electron number density (c) Variation of plasmon energy versus free electron number density (d) Variation of plasmon speed versus free electron number density.}
\end{figure}

The quasiparticle model of the electron gas in Eq. (\ref{eqp}) leads to the linearized stationary solution under conditions $\psi^0=1,\phi^0=0,\mu^0=\mu_0$ and separation of variable. The normalized system follows
\begin{subequations}\label{pf}
\begin{align}
&i\hbar\frac{{d\varphi(t)}}{{dt}} = \epsilon\varphi(t),\\
&\Delta \Psi ({\bf{r}}) + (\kappa\cdot\nabla) \Psi({\bf{r}}) + \Phi ({\bf{r}}) + 2E\Psi ({\bf{r}}) = 0,\\
&\Delta \Phi ({\bf{r}}) + (\kappa\cdot\nabla) \Phi({\bf{r}}) - \Psi ({\bf{r}}) = 0.
\end{align}
\end{subequations}
in which ${{\cal{N}}}({\bf{r}},t)=\psi({\bf{r}})\varphi(t)$ with $\Psi({\bf{r}})=\psi({\bf{r}})/\sqrt{n_0}$ and $\Phi({\bf{r}})=e\phi({\bf{r}})e^{i\theta}/2E_p$ are normalized quantities with $E_p=\sqrt{4\pi e^2 n_0/m}$ being the plasmon energy. Note that upon the linearization the phase of wavefunction $\Psi$ is lost, hence, the electrostatic potential in this system is considered imaginary in order to compensate this effect. Here, the normalized energy parameter $E=(\epsilon-\mu_0)/2E_p$ denotes the scaled total kinetic energy of the collective excitations with $\epsilon = \sum\limits_{j = 1}^N {{\epsilon_j}}$ being the energy eigenvalue of collective excitations with $\epsilon_j$ being the energy of $j$-th electron in the system. The normalization scheme leads to the space and time being also normalized with respect to the plasmon length $l_p=1/k_p$ with $k_p=\sqrt{2m E_p}/\hbar$ being the plasmon wavenumber and the inverse plasmon frequency, $\omega_p=E_p/\hbar$. Being in the plasmon unit scale requires all speeds to be normalized with respect to plasmon speed $v_p=\hbar k_p/m$ and temperatures with that of the plasmon $T_p=E_p/k_B$. The time dependent of stationary quasiparticle excitations is readily found to be $\varphi(t)=\exp(-i\omega t)$ with $\omega=\epsilon/\hbar$.

Fourier analysis of the system (\ref{pf}) leads to the generalized quasiparticle energy band of the form $E=(k^2+\kappa^2)/2+1/2(k^2+\kappa^2)$. In the absence of damping the well-known dispersion of collective excitation is retained which can also be written in a more useful form of $E=k_1^2/2+k_2^2/2$ where $k_1=\sqrt{E-\sqrt{E^2-1}}$ and $k_2=\sqrt{E-\sqrt{E^2+1}}$ denote the two distinct scale lengths known as the generalized de-Broglies wavenumbers characterizing wave-like and particle-like oscillations in the system and satisfy the complementarity relation $k_1 k_2=1$. 

Figure 1 depicts the generalized quasiparticle energy dispersion relation and the related plasmon parameter variations. Figure 1(a) shows the free electron band as dashed curve, the free quasiparticle band with thin solid curve and damped quasiparticle band as thick curve. The presence of collective excitation energy gap of $\Delta E=2E_p$ is apparent for quasiparticle bands. Both undamped and damped quasiparticle dispersions approach that of the free electron in the high wavenumber limit. The novel aspect of quasiparticle model is clearly reflected in this plot which captures both single particle and collective electron excitations in a single frame. Figure 1(b) shows the variations of plasmon scaling length in quasiparticle theory in nanometer unit. It varies from few tenth of a nanometer in typical metals up to tens of nanometers for electron gas in doped semiconductors. Figure 1(c) shows the variations of plasmon energy as the scaling unit of energy in this analysis and it varies from few electron volts in typical metals to relatively small values in semiconductors. Finally, Fig. 1(d) shows the plasmon speed variations with electron number density. It shows typical orders of $10^8$cm$/$s in typical metals as expected.

\section{Probability Current Density of Damped Quasiparticles}

Starting from the time-dependent damped 3D Schr\"{o}dinger-Poisson equation, we have \cite{akbdual}
\begin{subequations}\label{nsp}
\begin{align}
&2i\frac{{\partial {\cal {N}}({\bf{r}},t)}}{{\partial t}} =  - \Delta {\cal {N}}({\bf{r}},t) - 2(\kappa  \cdot \nabla ){\cal {N}}({\bf{r}},t) - \Phi ({\bf{r}}){\cal {N}}({\bf{r}},t) + \mu {\cal {N}}({\bf{r}},t),\\
&\Delta \Phi ({\bf{r}}) + 2(\kappa  \cdot \nabla )\Phi ({\bf{r}}) - {\cal {N}}({\bf{r}},t){N^*}({\bf{r}},t) = 0,
\end{align}
\end{subequations}
where the factor 2 again appears in (\ref{nsp}) because the time is normalized to $1/(2\omega_p)$. The continuity equation gives
\begin{equation}\label{con}
\frac{{\partial n}}{{\partial t}} =  \frac{{\partial {\cal {N}} ({\bf{r}},t){{\cal {N}} ^*}({\bf{r}},t)}}{{\partial t}} =  {\cal {N}} ({\bf{r}},t)\frac{{\partial {{\cal {N}} ^*}({\bf{r}},t)}}{{\partial t}} + {{\cal {N}} ^*}({\bf{r}},t)\frac{{\partial {\cal {N}} ({\bf{r}},t)}}{{\partial t}},
\end{equation}
where, the algebraic manipulation of time-dependent Schr\"{o}dinger-Poisson equation leads to 
\begin{subequations}\label{nsp1}
\begin{align}
&2i{{\cal {N}}^*}\frac{{\partial {\cal {N}}}}{{\partial t}} =  - {{\cal {N}}^*}\Delta {\cal {N}} - {{\cal {N}}^*}(\kappa  \cdot \nabla ){\cal {N}} - {{\cal {N}}^*}\Phi {\cal {N}} + {{\cal {N}}^*}\mu {\cal {N}}\\
&2i{\cal {N}}\frac{{\partial {{\cal {N}}^*}}}{{\partial t}} =  - {\cal {N}}\Delta {{\cal {N}}^*} - {\cal {N}}(\kappa  \cdot \nabla ){{\cal {N}}^*} - {\cal {N}}\Phi {{\cal {N}}^*} + {\cal {N}}\mu {{\cal {N}}^*}.
\end{align}
\end{subequations}
Combining the relations (\ref{con}) and (\ref{nsp1}), we find
\begin{equation}\label{con1}
-\frac{{\partial n}}{{\partial t}} = \frac{i}{2}\left( {{\cal {N}}\Delta {{\cal {N}}^*} - {{\cal {N}}^*}\Delta {\cal {N}}} \right) + i\left[ {{\cal {N}}(\kappa  \cdot \nabla ){{\cal {N}}^*} - {{\cal {N}}^*}(\kappa  \cdot \nabla ){\cal {N}}} \right] - \frac{i}{2}{\cal {N}}{{\cal {N}}^*}\left[ {\Phi  - {\Phi ^*}} \right].
\end{equation}
From the Poisson's relation, we obtain
\begin{equation}\label{poi}
{\cal {N}}{{\cal {N}}^*}\left( {\Phi  - {\Phi ^*}} \right) = \left( {\Phi \Delta {\Phi ^*} - {\Phi ^*}\Delta \Phi } \right) + 2i\left[ {\Phi (\kappa  \cdot \nabla ){\Phi ^*} - {\Phi ^*}(\kappa  \cdot \nabla )\Phi } \right],
\end{equation}
Now considering the generalized continuity equation $\partial n/\partial t + \nabla  \cdot {\bf{J}}_n = {S_n}$, where, ${\bf{J}}_n$ and $S_n$ denote the probability current density and the corresponding source term, we find that
\begin{equation}\label{jn}
{\bf{J}}_n({\bf{r}},t) = \frac{i}{2}\left[ {{\cal {N}} ({\bf{r}},t)\nabla {{\cal {N}} ^*}({\bf{r}},t) - {{\cal {N}} ^*}({\bf{r}},t)\nabla {\cal {N}} ({\bf{r}},t)} \right] - \frac{i}{2}\left[ {\Phi ({\bf{r}})\nabla {\Phi ^*}({\bf{r}}) - {\Phi ^*}({\bf{r}})\nabla \Phi ({\bf{r}})} \right].
\end{equation}
and
\begin{equation}\label{sn}
{S_n}({\bf{r}},t) =  - i\left[ {{\cal {N}}(\kappa  \cdot \nabla ){{\cal {N}}^*} - {{\cal {N}}^*}(\kappa  \cdot \nabla ){\cal {N}}} \right] + i\left[ {\Phi (\kappa  \cdot \nabla ){\Phi ^*} - {\Phi ^*}(\nabla  \cdot \kappa )\Phi } \right].
\end{equation}
Note that the source term vanishes for the undamped excitations.

\section{Generalized Energy Density of Collective Excitations}

In order to obtain the energy density, we use the standard definition for kinetic energy expected value of $\left\langle \epsilon  \right\rangle  = \int {{{\cal {N}}^*}{\cal {H}}{\cal {N}}d\tau }$ in which ${\cal {H}}$ is the Hamiltonian of the interacting electron system with the damping effect and $d\tau$ is the volume element. The Hamiltonian is given as
\begin{equation}\label{h}
{\cal {H}} = - \Delta  - 2(\kappa \cdot \nabla ) - \Phi ({\bf{r}}) + \mu.
\end{equation}
For quasiparticle excitations with scaled energy $E=\epsilon-\mu$ we have
\begin{equation}\label{E}
\left\langle E \right\rangle = - \int {{{\cal {N}}^*}({\bf{r}},t)\Delta {\cal {N}}({\bf{r}},t)d\tau }  - 2\int {{{\cal {N}}^*}({\bf{r}},t)(\kappa  \cdot \nabla ){\cal {N}}({\bf{r}},t)d\tau }  - \int {{{\cal {N}}^*}({\bf{r}},t)\Phi ({\bf{r}}){\cal {N}}({\bf{r}},t)d\tau }.
\end{equation}
It follows that the energy density of the system is given by
\begin{equation}\label{Ed}
{\rho _E} =  - {{\cal {N}}^*}({\bf{r}},t)\Delta {\cal {N}}({\bf{r}},t) - 2{{\cal {N}}^*}({\bf{r}},t)(\nabla  \cdot \kappa ){\cal {N}}({\bf{r}},t) - {{\cal {N}}^*}({\bf{r}},t)\Phi ({\bf{r}}){\cal {N}}({\bf{r}},t).
\end{equation}
From the damped Poisson's relation we have
\begin{equation}\label{Pd}
\Phi \Delta {\Phi ^*} + 2\Phi (\kappa  \cdot \nabla ){\Phi ^*} - {{\cal {N}}^*}\Phi {\cal {N}} = 0,
\end{equation}
which consequently leads to
\begin{equation}\label{Pd2}
{\rho _E} =  - {{\cal {N}}^*}\Delta {\cal {N}} - 2{{\cal {N}}^*}(\kappa  \cdot \nabla ){\cal {N}} - \Phi \Delta {\Phi ^*} + 2\Phi \left( {\kappa \cdot \nabla } \right){\Phi ^*}.
\end{equation}
Now using the identities ${{\cal {N}} ^*}\Delta {\cal {N}}  = \nabla \cdot\left( {{{\cal {N}} ^*}\nabla {\cal {N}} } \right) - \nabla {{\cal {N}} ^*}\cdot\nabla {\cal {N}}$ and $\Phi \Delta {\Phi ^*} = \nabla \cdot\left( {\Phi \nabla {\Phi ^*}} \right) - \nabla {\Phi ^*}\cdot\nabla \Phi$ we arrive at the following expression for the energy density
\begin{equation}\label{Pd3}
{\rho _E} = \nabla {{\cal {N}}^*} \cdot \nabla {\cal {N}} + \nabla {\Phi ^*} \cdot \nabla \Phi  - \nabla  \cdot \left( {{{\cal {N}}^*}\nabla {\cal {N}}} \right) - \nabla  \cdot \left( {\Phi \nabla {\Phi ^*}} \right) - 2{{\cal {N}}^*}\left( {\kappa  \cdot \nabla } \right){\cal {N}} - 2\Phi \left( {\kappa  \cdot \nabla } \right){\Phi ^*}.
\end{equation}
The first term in (\ref{Pd3}) is the normal kinetic energy density \cite{mita}. However, the second term represents the electrostatic potential energy density of the system since ${\cal{E}}=-\nabla\Phi$ represents the normalized local electric field in the electron gas. The remaining terms in (\ref{Pd3}) are cast in terms of the $\Re$ (real) and $\Im$ (imaginary) components given as follows 
\begin{subequations}\label{reim}
\begin{align}
&\Re \left[ {{{\cal {N}}^*}\left( {\kappa  \cdot \nabla } \right){\cal {N}}} \right] = \frac{1}{2}\left[ {{{\cal {N}}^*}\left( {\kappa  \cdot \nabla } \right){\cal {N}} + {\cal {N}}\left( {\kappa  \cdot \nabla } \right){{\cal {N}}^*}} \right],\\
&\Im \left[ {{{\cal {N}}^*}\left( {\kappa  \cdot \nabla } \right){\cal {N}}} \right] = \frac{i}{2}\left[ {{{\cal {N}}}\left( {\kappa  \cdot \nabla } \right){\cal {N}^*} - {\cal {N}^*}\left( {\kappa  \cdot \nabla } \right){{\cal {N}}}} \right],\\
&\Re \left[ {\Phi \left( {\kappa  \cdot \nabla } \right){\Phi ^*}} \right] = \frac{1}{2}\left[ {\Phi \left( {\kappa  \cdot \nabla } \right){\Phi ^*} + {\Phi ^*}\left( {\kappa  \cdot \nabla } \right)\Phi } \right],\\
&\Im \left[ {\Phi \left( {\kappa  \cdot \nabla } \right){\Phi ^*}} \right] = \frac{i}{2}\left[ {{\Phi ^*}\left( {\kappa  \cdot \nabla } \right)\Phi  - \Phi \left( {\kappa  \cdot \nabla } \right){\Phi ^*}} \right].
\end{align}
\end{subequations}
This results in a compact form
\begin{equation}\label{Pd4}
{\rho _E} = \nabla {{\cal {N}}^*} \cdot \nabla {\cal {N}} + \nabla {\Phi ^*} \cdot \nabla \Phi  + \nabla \cdot \left( {{{\bf{J}}_\rho } - i{{\bf{J}}_n}} \right) - \left( {{S_\rho } - i{S_n}} \right).
\end{equation}
where ${\bf{J}}_\rho= -\frac{1}{2}\nabla \rho$ is the quantum diffusion current density with $\rho = {\cal {N}}{\cal {N}}^* + \Phi \Phi^*$ being the generalized density (energy-density) of the system. The continuity equation for diffusion current is $\partial \rho/\partial t + \nabla \cdot {\bf{J}}_\rho = S_\rho$ with the diffusion source term given as $S_\rho=(\kappa \cdot \nabla)\rho$.
The final form of the energy density is given as
\begin{equation}\label{Pdf}
{\rho _E} = \left( {\nabla {{\cal {N}}^*} \cdot \nabla {\cal {N}} + \nabla {\Phi ^*} \cdot \nabla \Phi } \right) - \frac{{\partial \rho }}{{\partial t}} + i\frac{{\partial n}}{{\partial t}},
\end{equation}
where $n={\cal {N}}{\cal {N}}^*$ is the normal density. Note that the appearance of imaginary term in the generalized energy density relation (\ref{Pdf}) represents the existence of damping or growing quantum instability. Equation (\ref{Pdf}) can be regarded as the generalized form of the equation given in \cite{mita}.  

\section{One-Dimensional Free Quasiparticle Excitations}

\begin{figure}[ptb]\label{Figure2}
\includegraphics[scale=0.6]{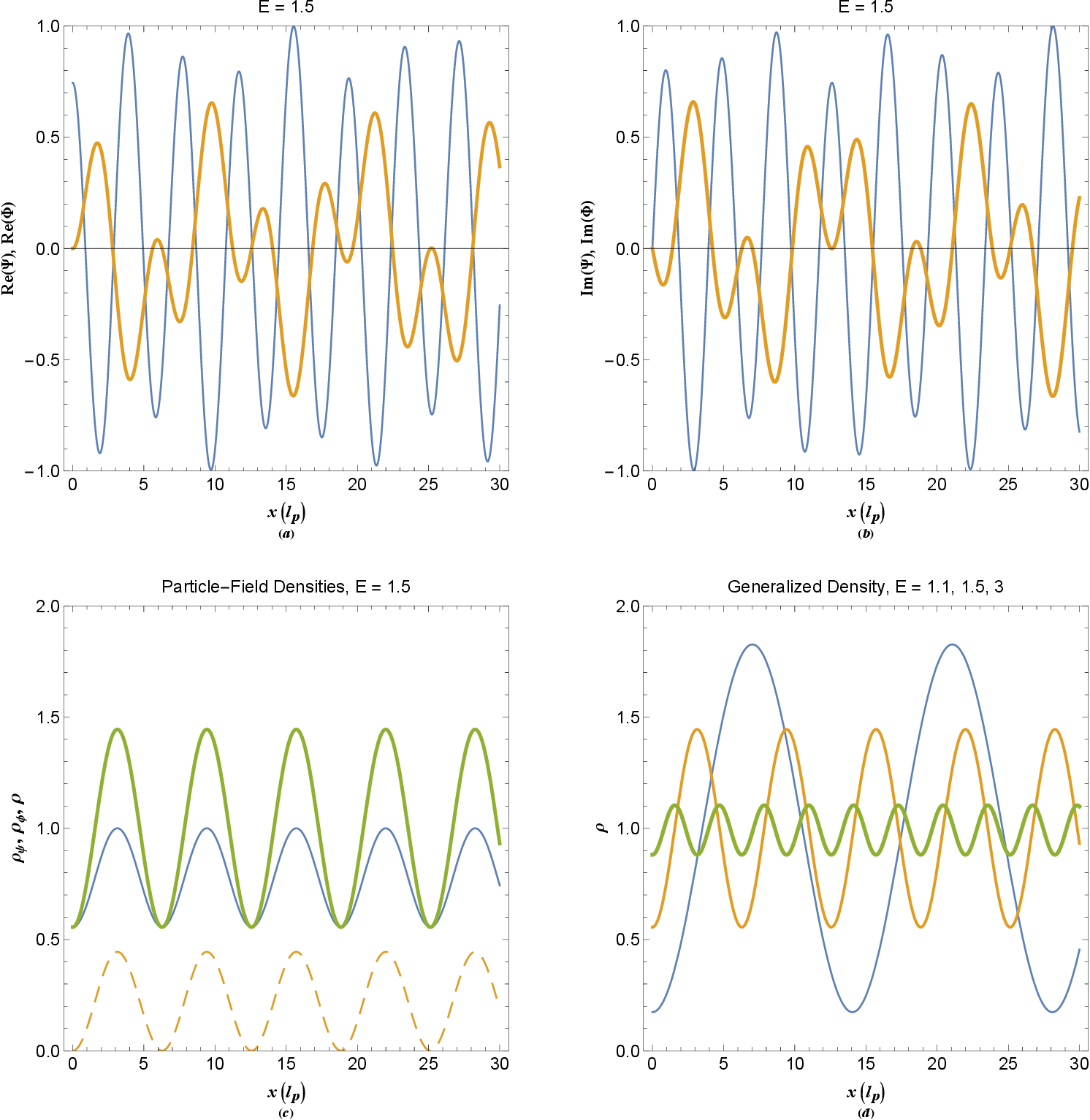}\caption{(a) Real part of free quasiparticle wavefunction (thin curve) and electrostatic energy (thick curve) profiles. (b) Imaginary part of free quasiparticle wavefunction (thin curve) and electrostatic energy (thick curve) profiles. (c) The generalized density due to particle (thin solid curve), field (dashed curve) and total (thick curve). (d) The total generalized density for different quasiparticle energy orbital. Increase in thickness in (d) reflects the increase in varied parameter in this plot.}
\end{figure}

We now apply the findings of the proceeding section to the free quasiparticle excitations in the arbitrary degenerate electron gas. The one-dimensional solution of type ${\cal{N}}=\Psi(x)\exp(-i\omega t)$ ($\omega=\epsilon/\hbar$) to (\ref{nsp}) with $\kappa=0$ and initial values $\Psi'_0=\Phi'_0=0$ reads \cite{akbdual}
\begin{equation}\label{wf}
\left[ {\begin{array}{*{20}{c}}
{{\Phi}(x)}\\
{{\Psi}(x)}
\end{array}} \right] = \frac{A_\rho}{{2\alpha }}\left[ {\begin{array}{*{20}{c}}
{{\Psi _0} + k_2^2{\Phi _0}}&{ - \left( {{\Psi _0} + k_1^2{\Phi _0}} \right)}\\
{ - \left( {{\Phi _0} + k_1^2{\Psi _0}} \right)}&{{\Phi _0} + k_2^2{\Psi _0}}
\end{array}} \right]\left[ {\begin{array}{*{20}{c}}
{\exp (i{k_1}x)}\\
{\exp (i{k_2}x)}
\end{array}} \right],
\end{equation}
where $A_\rho$ is the normalization factor to ensure the consistency with the equilibrium state, ${k_1} = \sqrt {E - \alpha }$, ${k_2} = \sqrt {E + \alpha }$ with $\alpha  = \sqrt {{E^2} - 1}$ admitting a complementarity-like relation between wave-like and particle-like length scales ($k_1 k_2=1$). The initial values are set $\Psi_0=1$ and $\Phi_0=0$ throughout the analysis, as is expected for positions far from perturbations. The free quasiparticle theory is useful in modeling of linear collective excitations in a dense electron beam or other unbounded electron fluid. The excitations which may be caused by external stimuli such as electromagnetic interactions. In this case the particle, potential and generalized densities are given respectively, as
\begin{subequations}\label{rho0}
\begin{align}
&{\rho _\psi } = \frac{A_\rho^2}{{4{\alpha ^2}}}\left\{ {k_1^4 + k_2^4 - 2\cos \left[ {\left( {{k_1} - {k_2}} \right)x} \right]} \right\},\\
&{\rho _\phi } = \frac{A_\rho^2}{{{\alpha ^2}}}\sin {\left[ {\frac{1}{2}\left( {{k_1} - {k_2}} \right)x} \right]^2},\\
&\rho  = \rho_\psi + \rho_\phi = \frac{A_\rho^2}{{4{\alpha ^2}}}\left\{ {2 + k_1^4 + k_2^4 - 4\cos \left[ {\left( {{k_1} - {k_2}} \right)x} \right]} \right\},
\end{align}
\end{subequations}
where the normalization factor $A_\rho(E)=2{\alpha}/\sqrt{2 + k_1^4 + k_2^4}$ equals the invariant quantity of the system, $\rho_\psi-\rho_\phi=1-1/E^2$, for given quasiparticle orbital energy, $E$. This must be compared to the case of free electron wavefunction which is nonnormalizable.

Figure 2 depicts normalized wavefucntion (thin curves) and electrostatic energy (thick curves) profiles of free quasiparticle excitations along with generalized density variations. Figure 2(a) and 2(b) shows the real and imaginary parts at the quasiparticle orbital $E=1.5$ indicating the dual-tone nature of oscillations. The generalized density $\rho=\Psi^*\Psi+\Phi^*\Phi$ (thick curve), with $\rho_\psi=\Psi^*\Psi$ (thin curve) and $\rho_\phi=\Phi^*\Phi$ (dashed curve) profiles is depicted in Fig. 2(c). Note that the quantity $\Psi^*\Psi-\Phi^*\Phi$ is space invariant in the system. The generalized density oscillates around the equilibrium value of $\rho_0=\Psi_0^2+\Phi_0^2$. Figure 2(d) shows the generalized density profiles at different quasiparticle orbital. The amplitude of oscillations is increased/decreased as its wavelength is increased/decreased which is a charcteristic behavior of the quasiparticle dispersion, shown in Fig. 1(a).   

\begin{figure}[ptb]\label{Figure3}
\includegraphics[scale=0.6]{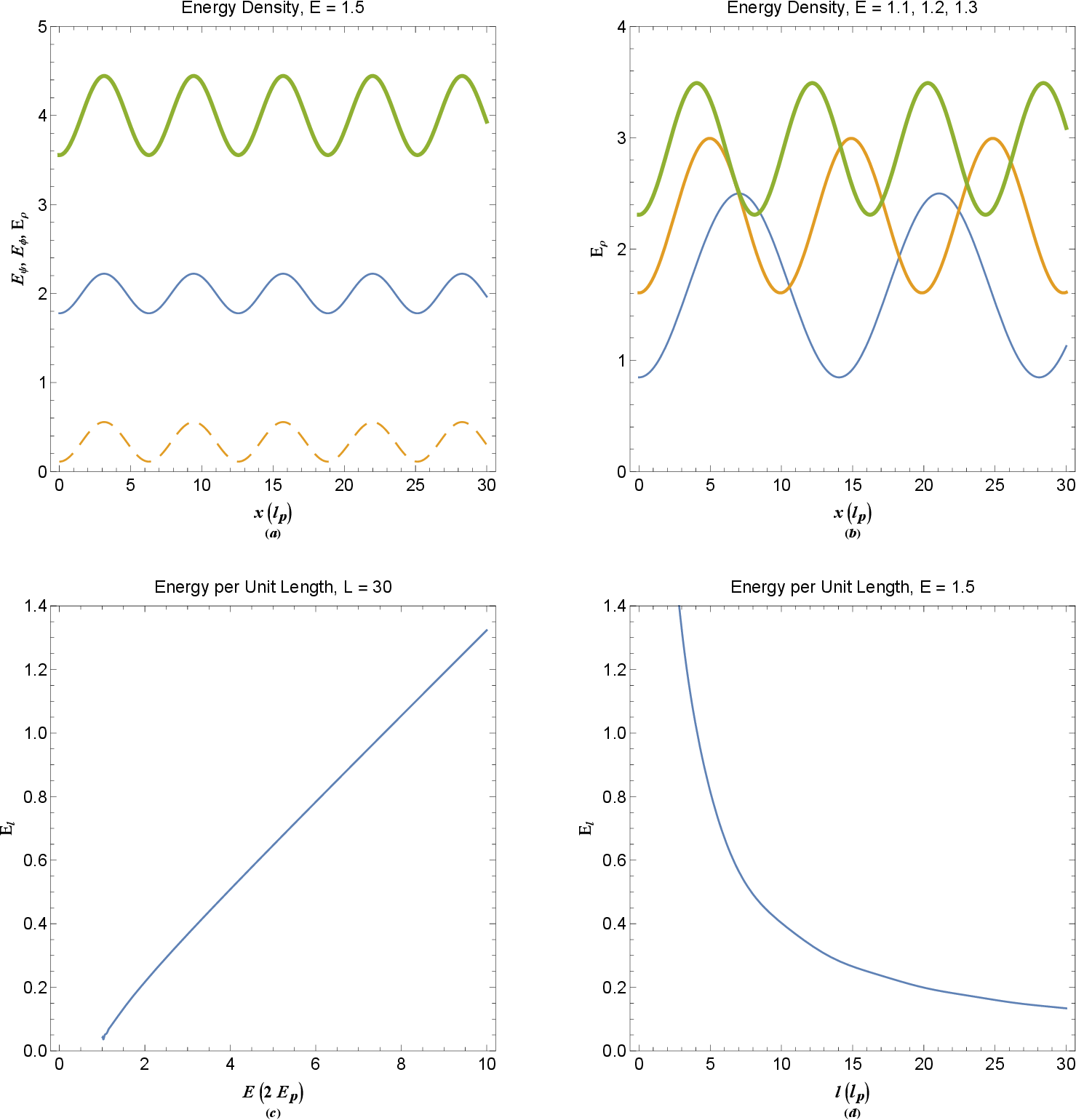}\caption{(a) Energy density of free quasiparticle excitations due to kinetic (thin solid curve), potential (dashed curve) and total (thick curve). (b) Total energy density for different orbital energies with curve thickness increasing with increase in the energy. (c) The expected energy per unit length variation with quasiparticle orbital energy. (d) The expected energy per unit length varied with length.}
\end{figure}

The energy densities corresponding to kinetic, potential and generalized densities are respectively given as 
\begin{subequations}\label{E0}
\begin{align}
&{E_\psi } = \frac{{d{\Psi ^*}}}{{dx}}\frac{{d\Psi }}{{dx}} = \frac{A_\rho^2}{{4{\alpha ^2}}}\left\{ {k_1^6 + k_2^6 - 2\cos \left[ {\left( {{k_1} - {k_2}} \right)x} \right]} \right\},\\
&{E_\phi } = \frac{{d{\Phi ^*}}}{{dx}}\frac{{d\Phi }}{{dx}} = \frac{A_\rho^2}{{4{\alpha ^2}}}\left\{ {k_1^2 + k_2^2 - 2\cos \left[ {\left( {{k_1} - {k_2}} \right)x} \right]} \right\},\\
&{E_\rho } = {E_\psi } + {E_\phi } = \frac{A_\rho^2}{{2{\alpha ^2}}}\left\{ {k_1^6 + k_2^6 - 2\cos \left[ {\left( {{k_1} - {k_2}} \right)x} \right]} \right\}.
\end{align}
\end{subequations}
Note that $E_\rho=2E_\psi$.

Figure 3 depicts the variations in energy density of free quasiparticle excitations in the electron gas at given energy orbital. Figure 3(a) reveals that the oscillation wavelength in kinetic ($E_\psi$) and potential ($E_\phi$) energy densities is the same, so that, the oscillate in-phase together. It is because the kinetic energy comes from the particle aspect of the gas where as the potential energy from the charge which are indeed attached to each other in the electron gas. Figure 3(b) shows that increase in orbital energy leads to increase the level but decrease in wavelength of the total energy density ($E_\rho=E_\psi+E_\phi$). The expected quasiparticle energy per unit length is shown in terms of orbital energy in Fig. 3(c) for $L=30$ in plasmon length unit. It is shown that this quantity increase with orbital energy, monotonically. On the other hand, Fig. 3(d) shows that for the orbital energy $E=1.5$ (in twice plasmon energy unit) the expected energy per unit length decreases with increase in length $L$.
     
\begin{figure}[ptb]\label{Figure4}
\includegraphics[scale=0.6]{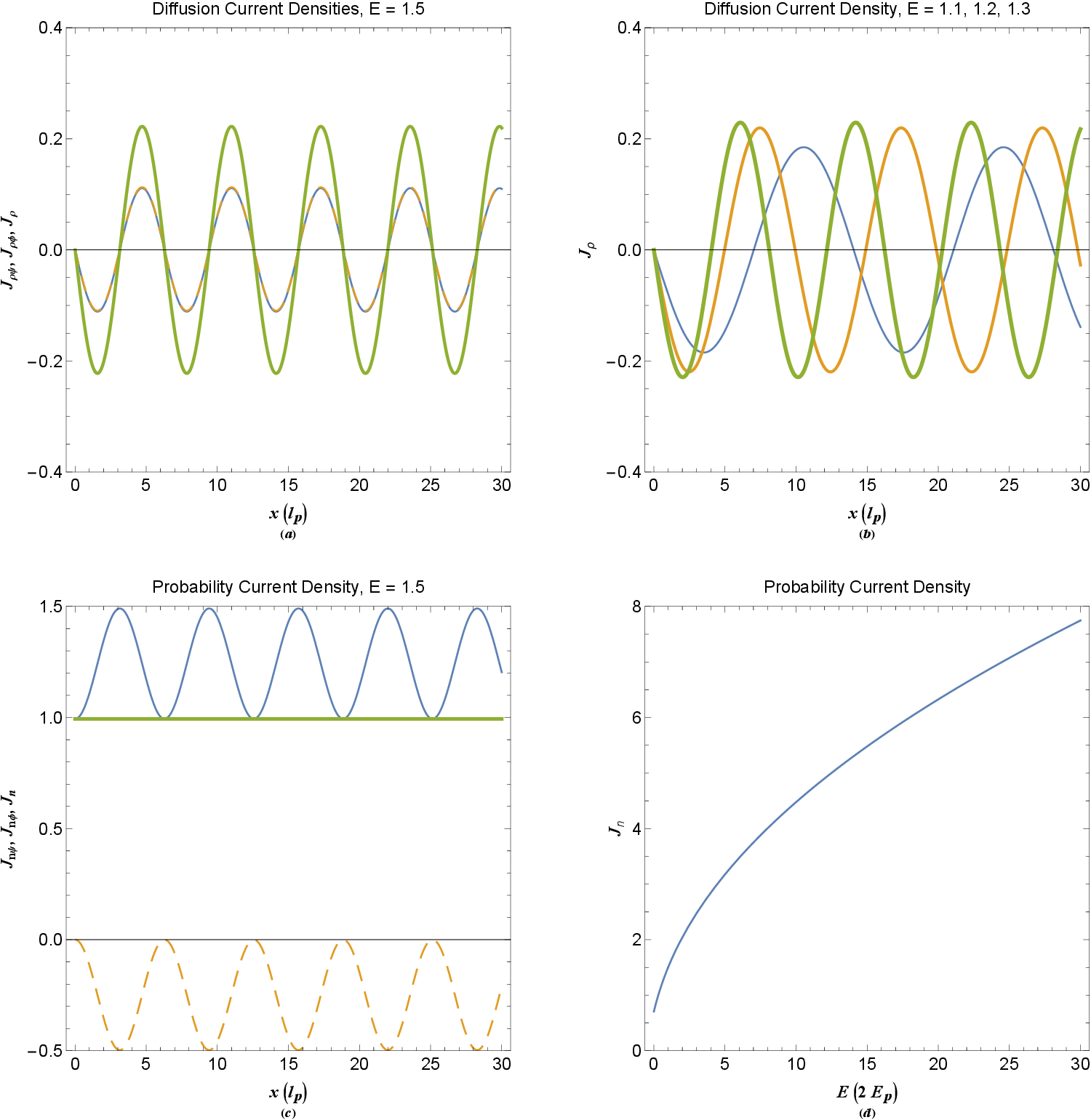}\caption{(a) Diffusion current densities for free quasiparticle excitations due to kinetic (thin solid curve), potential (dashed curve) and total (thick curve). (b) Total diffusion current density for different orbital energies with curve thickness increasing with increase in the energy. (c) probability current densities for free quasiparticle excitations due to kinetic (thin solid curve), potential (dashed curve) and total (thick curve). (d) Variation of total probability current with quasiparticle orbital energy.}
\end{figure}

On the other hand, the diffusion current densities can be written as 
\begin{equation}\label{dj}
{J_{\rho \psi }} = {J_{\rho \phi }} = \frac{A_\rho^2}{{4{\alpha ^2}}}\left( {{k_2} - {k_1}} \right)\sin \left[ {\left( {{k_1} - {k_2}} \right)x} \right],
\end{equation}
and the total diffusion current reads ${J_{\rho }}=2{J_{\rho \psi }}$. Moreover, the diffusion current flux, given that the diffusion source is zero, reads
\begin{equation}\label{df}
-\frac{dJ_\rho}{dx} =  - \frac{A_\rho^2}{{2{\alpha ^2}}}{\left( {{k_1} - {k_2}} \right)^2}\cos \left[ {\left( {{k_1} - {k_2}} \right)x} \right],
\end{equation}
which indicates the local density flow in or out of the point.  

The space variations in diffusion current densities are depicted in Fig. 4. Figure 4(a) reveals that the current densities at orbital $E=1.5$, contributed from kinetic ($J_{\rho\psi}$) and potential ($J_{\rho\phi}$) energies, amount to the same value in all space. It is remarked from Fig. 4(b) that magnitude/wavelength of the diffusion current densities increase/decrease with the orbital quasiparticle energy. The probability current density variations are shown in Figs. 4(c) and 4(d). It is shown in Fig. 4(c) that contributions from kinetic ($J_{n\psi}$) and potential ($J_{n\phi}$) vary out of phase and amounts to a constant value of $J_n=J_{n\psi}+J_{n\phi}=(k_1^5+k_2^5-k_1-k_2)/4 E^2$ for the given orbital energy $E$. Variation of probability current density with orbital energy is depicted in Fig. 4(d) showing a monotonic increase in this quantity.  

\begin{figure}[ptb]\label{Figure5}
\includegraphics[scale=0.6]{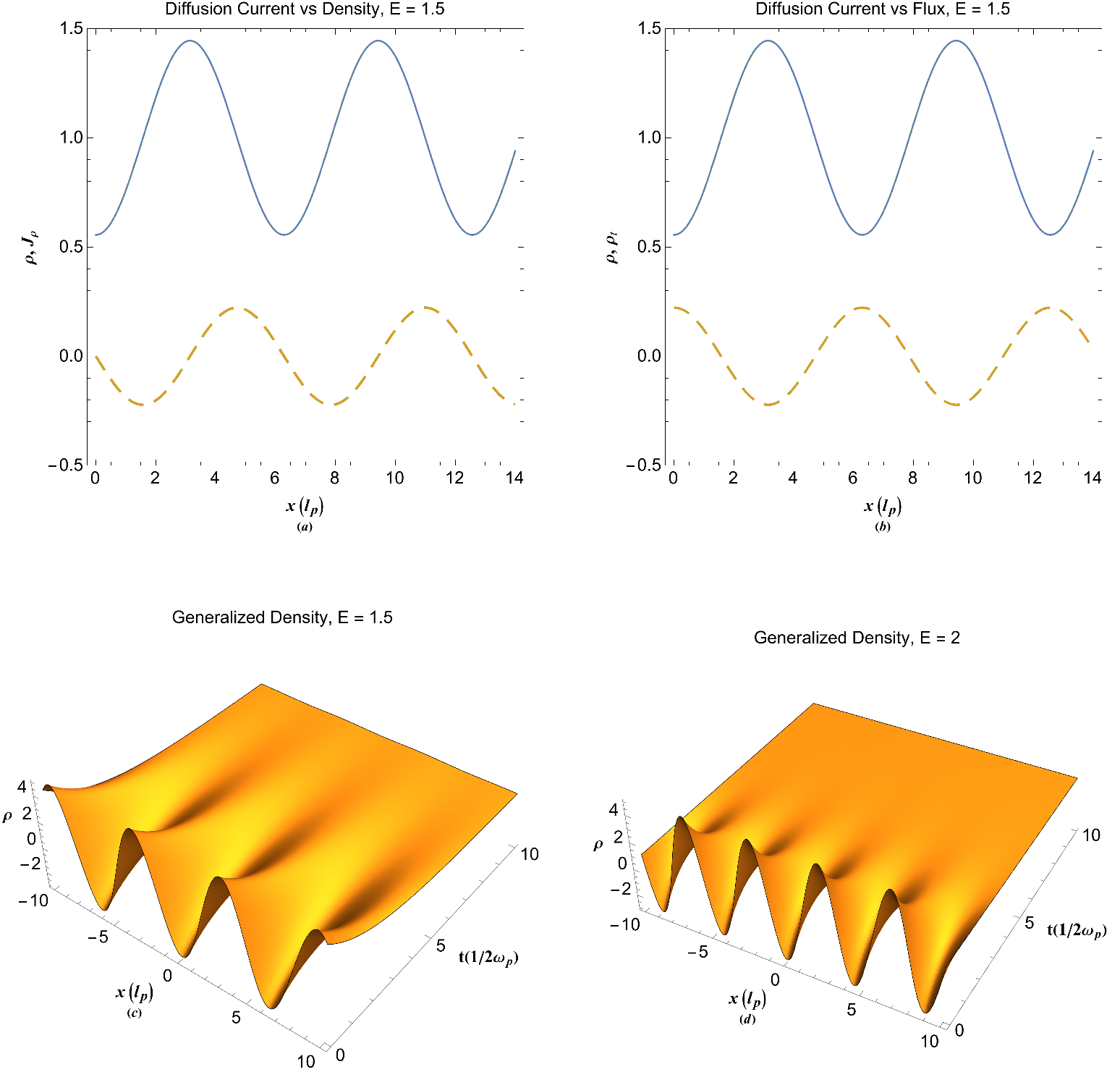}\caption{(a) Variation of diffusion current density (dashed curve) and generalized density (solid curve). (b) Variation of diffusion density flux (dashed curve) and generalized density (solid curve). (c) Diffusion of excited free quasiparticle state from orbital with energy $E=1.5$. (d) Diffusion of excited free quasiparticle state from orbital with energy $E=2$.}
\end{figure}

The probability current densities are given as  
\begin{subequations}\label{pj}
\begin{align}
&{J_{n \psi }} = \frac{A_\rho^2}{{4{\alpha ^2}}}\left\{ {k_1^5 + k_2^5 - \left( {{k_1} + {k_2}} \right)\cos \left[ {\left( {{k_1} - {k_2}} \right)x} \right]} \right\},\\
&{J_{n \phi }} = \frac{{{A_\rho^2 }}}{{4{\alpha ^2}}}\left( {{k_1} + {k_2}} \right)\left\{ {\cos \left[ {\left( {{k_1} - {k_2}} \right)x} \right] - 1} \right\},
\end{align}
\end{subequations}
and the total probability current reads ${J_{n}}=A_\rho^2(k_1^5 + k_2^5 - {k_1} - {k_2})/4{\alpha ^2}$ which is constant. Also, the probability flux $\partial n/\partial t$ is zero for all given quasiparticle energy orbital. The energy per unit length of the 1D electron gas at given orbital $E$ is
\begin{equation}\label{el}
{E_l} = \frac{1}{l}\int\limits_0^l {{E_\rho }dx}  = \frac{2/l}{{2 + k_1^4 + k_2^4}}\left\{ {k_1^6 + k_2^6 - \frac{{2\sin \left[ {\left( {{k_1} - {k_2}} \right)l} \right]}}{{\left( {{k_1} - {k_2}} \right)l}}} \right\}.
\end{equation} 
The time evolution the generalized density in the system is given by the diffusion-like equation $\partial \rho(x,t)/\partial t=(1/2)\partial^2\rho(x,t)/\partial x^2$ with the initial condition $\rho(x,0)$ given by Eq. (\ref{rho0})c. The time dependent solution then reads
\begin{equation}\label{rhot0}
\rho (x,t) = 1 - 4{{{e}}^{ - \frac{1}{2}{{\left( {{k_1} - {k_2}} \right)}^2}t}}\cos \left[ {\left( {{k_1} - {k_2}} \right)x} \right].
\end{equation}
Note that $\rho(x,\infty)=1$, as is dictated by the normalization value. The characteristic quantum diffusion or the perturbation damping time is $\eta=2/(k_1-k_2)^2=1/(E-1)$ inversely proportional to the quasiparticle orbital energy. This is to say that collective excitations with higher energy diffuse faster towards the equilibrium state. Note also that the damping time diverges for $E=1$ which corresponds to quantum beating or ground-state quasiparticle orbital where $k_1=k_2$.

Figure 5 shows the density flux due to quantum diffusion in the electron gas. Figure 5(a) shows the diffusion current density (dashed curve) and generalized density (solid curve) in the same plot. The diffusion current varies sinusoidally according to the generalized density distribution in the electron gas. The governing continuity equation for this variations is $\partial \rho(x,t)/\partial t+\partial J_{\rho}(x,t)/\partial x=0$. The variation in diffusion density flux $\partial \rho(x,t)/\partial t$ with $\rho$ is depicted in Fig. 5(b). It is clearly remarked that the flux is inward/outward (positive/negative) on low/high density regions trying to retain the uniformity of the generalized density distribution in the electron gas. This effect is simulated in a 3D plot in Figs. 5(c) and 5(d) for different values of orbital energy. They clearly depict the quantum diffusion effect on collective excited states in the system. The later effect predicts a unstable character of all excited quasiparticle states except the ground state level for which the decay time becomes infinite. It is therefore essential for a long lived collective excitations to reside as close to the quantum beating point for which the exciting photon energy is $\hbar\omega\simeq 2E_p$.      

\section{Collective Electron Excitations in a 1D Box}

\begin{figure}[ptb]\label{Figure6}
\includegraphics[scale=0.6]{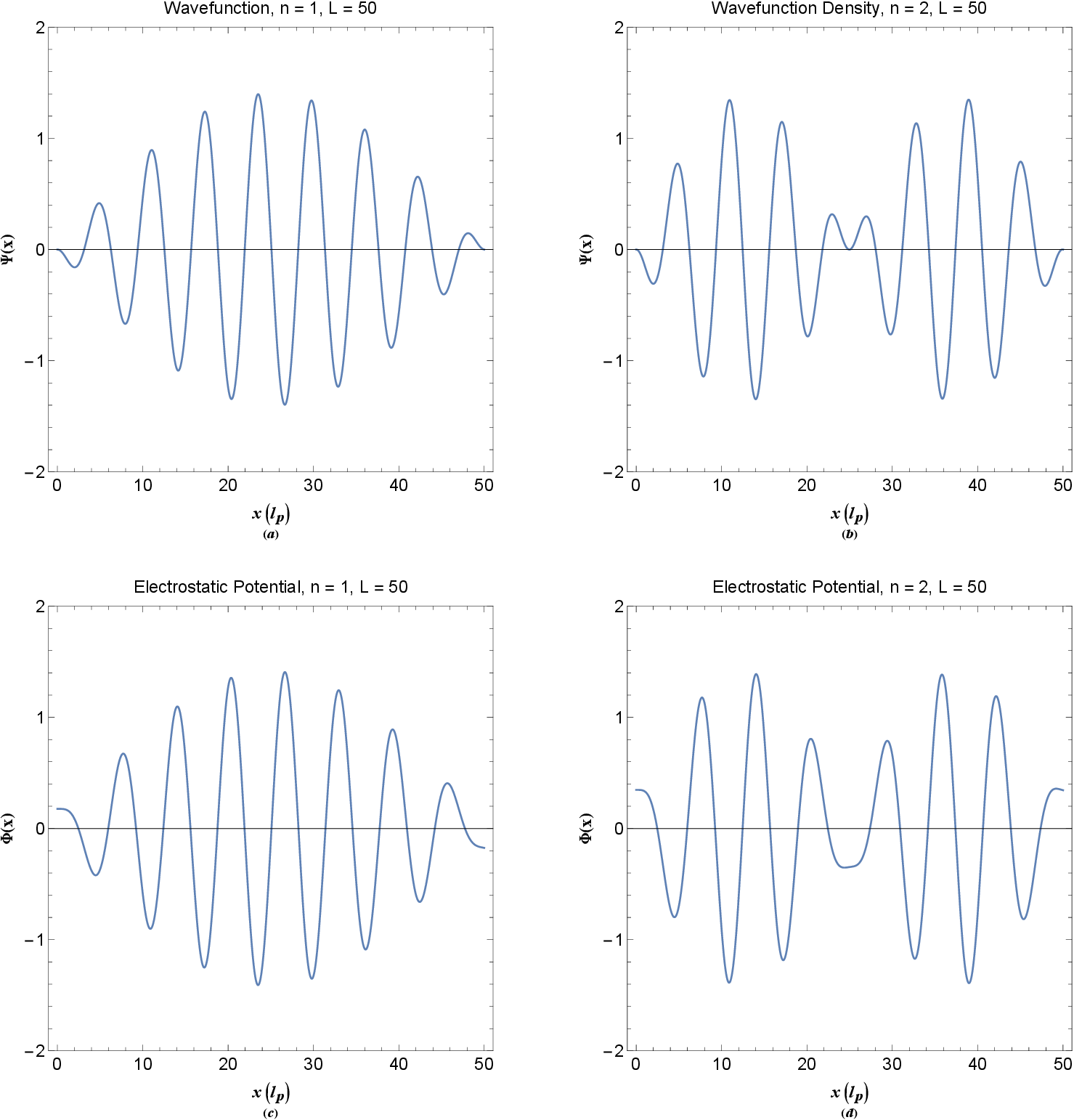}\caption{(a) Wavefunction of quasiparticle in a box at ground state. (b) Wavefunction of quasiparticle in a box at first excited state. (c) Electrostatic energy of quasiparticle in a box at ground state. (d) Electrostatic energy of quasiparticle in a box at first excited state.}
\end{figure}

We now turn into the solution of undamped quasiparticle excitations in one-dimensional infinite square-well of length $L$. The solution, with previously taken  initial values, is given by \cite{akbquant}
\begin{equation}\label{qbox}
\left[ {\begin{array}{*{20}{c}}
{\Phi (x)}\\
{\Psi (x)}
\end{array}} \right] = \frac{{{A_\rho }}}{{2\alpha }}\left[ {\begin{array}{*{20}{c}}
{k_2^2}&{ - k_1^2}\\
{ - 1}&1
\end{array}} \right]\left[ {\begin{array}{*{20}{c}}
{\cos ({k_1}x)}\\
{\cos ({k_2}x)}
\end{array}} \right],
\end{equation}
where $A$ is a normalization factor satisfying the relation $\rho(x,t)|_{t\to\infty}=1$ given as
\begin{equation}\label{norm}
A_\rho\left( {L,n} \right) = \frac{{2\sqrt 2 \alpha }}{{\sqrt {2 + k_1^4 + k_2^4} }} .
\end{equation}
The quasiparticle orbital in this case are quantized with the energy eigenvalues given by $E_n=(L^2 + 2 n^2 \pi^2)/L^2$.

The generalized density profiles for $L=40$ and $n=1,2$ is depicted in Fig. 7. The contribution from particle probability ($\rho_\psi$) is shown in Fig. 7(a). This contribution clearly vanished at the wall locations. The contribution from charge probability ($\rho_\phi$) is depicted in Fig. 7(b) showing a sinilar profile as in Fig. 7(a), but, with non-vanishing values at the walls. The total generalized density $\rho=\rho_\psi+\rho_\phi$ is depicted for $n=1,2$ in Figs. 7(c) and 7(d). These plots clearly indicate the characteristic features of quantum interference effect due to single electron and collective oscillations in the electron gas. 

Figure 6 depicts the normalized wavefunction and corresponding electrostatic energy distributions for a quasiparticle state in an infinite square well of length $L=50$ in the quantized levels $n=1,2$. It shows the characteristic wavefucntion showing variation both due to single electron as well as collective oscillations in the box. note that in this case, as compared to the elementary problem of particle in a box, the fine structure density oscillation is due to the single particle excitations. Also note that while the wavefucntion vanishes at the potential wall the electrostatic energy remains finite at this locations due to finite probability of charge close to the confining walls.

\begin{figure}[ptb]\label{Figure7}
\includegraphics[scale=0.6]{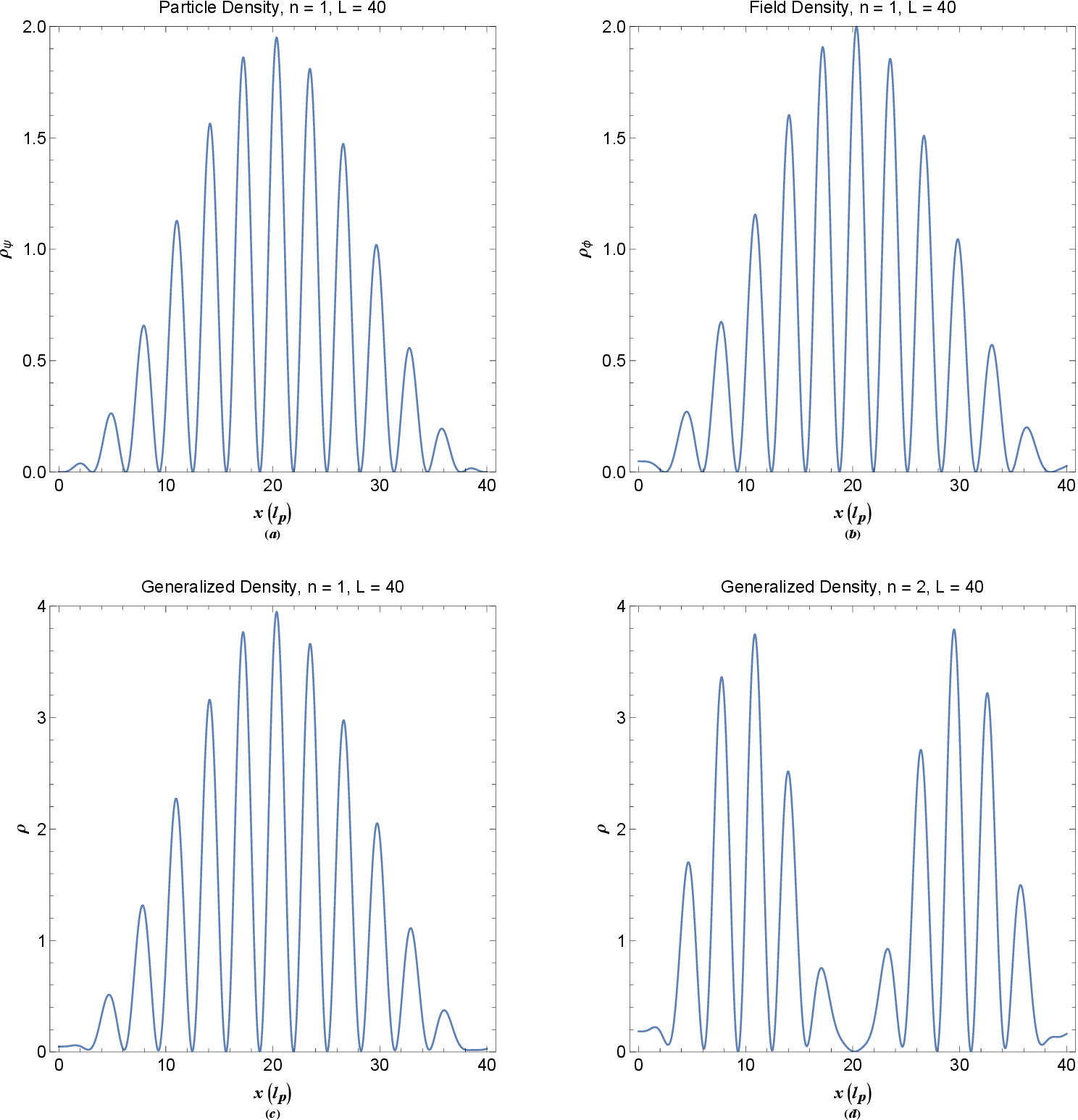}\caption{(a) Generalized density due to particle of quasiparticle in a box at ground state. (b) Generalized density due to charge of quasiparticle in a box at first excited state. (c) Total generalized density of quasiparticle in a box at ground state. (d) Total generalized density of quasiparticle in a box at first excited state.}
\end{figure}

The generalized density of quasiparticles in this case reads
\begin{equation}\label{rho1}
\rho \left( {L,n} \right) = \frac{{{A^2}}}{{4{\alpha ^2}}}\left[ {\left( {1 + k_2^4} \right){{\cos }^2}\left( {{k_1}x} \right) - 4\cos \left( {{k_1}x} \right)\cos \left( {{k_2}x} \right) + \left( {1 + k_1^4} \right){{\cos }^2}\left( {{k_2}x} \right)} \right].
\end{equation}
Note that the generalized density does not vanish at the boundary of potential well, since, it includes the potential squared values by the definition. 

\begin{figure}[ptb]\label{Figure8}
\includegraphics[scale=0.6]{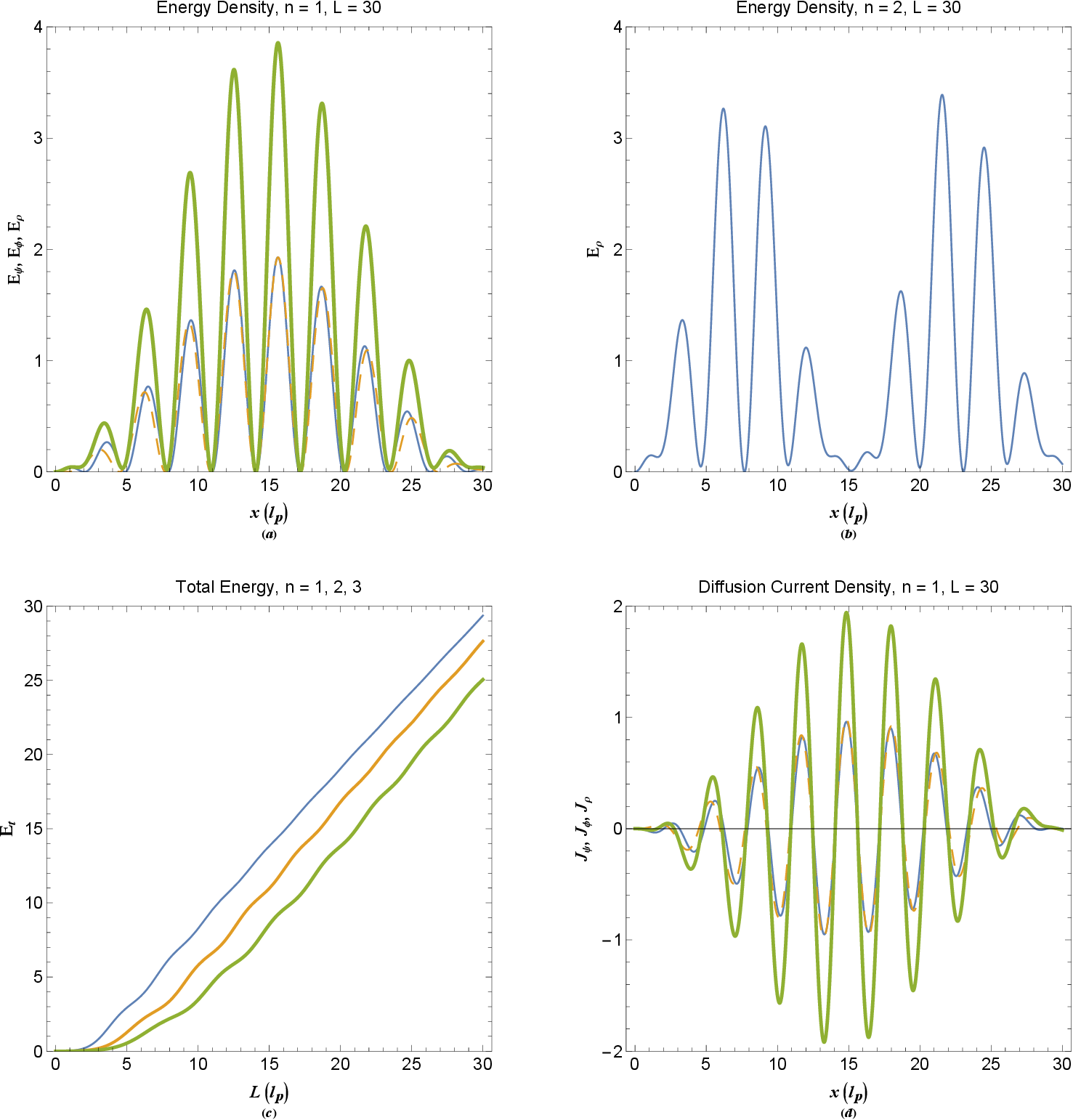}\caption{(a) Energy density of quasiparticle in a box kinetic (thin solid curve), potential (dashed curve) energies and total (thick curve). (b) Total energy density of quasiparticle in a box at first excited state. (c) Variation of normalized expected energy inside the box for quantum number values $n=1,2,3$ with resoect to the box length. Increase in curve thickness indicates the increase in varied parameter above the panel. (d) Variation of diffusion current density of quasiparticle in a box due to kinetic (thin solid curve), potential (dashed curve) energies and total diffusion current density (thick curve).}
\end{figure}

On the other hand, the energy density distributions due to particle and field presence are given as
\begin{subequations}\label{E1}
\begin{align}
&{E_\psi } = \frac{{d{\Psi ^*}}}{{dx}}\frac{{d\Psi }}{{dx}} = \frac{{{A_\rho^2}}}{{4{\alpha ^2}}}{\left[ {{k_1}\sin \left( {{k_1}x} \right) - {k_2}\sin \left( {{k_2}x} \right)} \right]^2},\\
&{E_\phi } = \frac{{d{\Phi ^*}}}{{dx}}\frac{{d\Phi }}{{dx}} = \frac{{{A_\rho^2}}}{{4{\alpha ^2}}}{\left[ {{k_2}\sin \left( {{k_1}x} \right) - {k_1}\sin \left( {{k_2}x} \right)} \right]^2},\\
&{E_\rho } = {E_\psi } + {E_\phi } = \frac{{{A_\rho^2}}}{{4{\alpha ^2}}}\left\{ {{{\left[ {{k_1}\sin \left( {{k_1}x} \right) - {k_2}\sin \left( {{k_2}x} \right)} \right]}^2} + {{\left[ {{k_2}\sin \left( {{k_1}x} \right) - {k_1}\sin \left( {{k_2}x} \right)} \right]}^2}} \right\}.
\end{align}
\end{subequations} 
Also the total energy inside the potential well is given as the orbital energy, width of the confining box and the quantum number, as
\begin{subequations}\label{El1}
\begin{align}
&{E_l} = \int\limits_0^L {{E_\rho }dx}  = \frac{{{A_\rho^2}}}{{16{\alpha ^2}}}\left\{ {{k_1}\left( {1 + k_2^4} \right)\left[ {2{k_1}L - \sin \left( {2{k_1}L} \right)} \right] + {k_2}\left( {1 + k_1^4} \right)\left[ {2{k_2}L - \sin \left( {2{k_2}L} \right)} \right]} \right.\\
&{ + \frac{{8\left[ {{k_1}\sin \left( {2{k_2}L} \right) - {k_2}\sin \left( {2{k_1}L} \right)} \right]}}{{k_1^2 - k_2^2}}}.
\end{align}
\end{subequations} 

Figure 8(a) shows the energy density distribution of quasiparticle in a box. While the contributions from kinetic (thin solid curve) and potential (dashed curve) energies amount approximately to the same level, there are variances when approaching the walls. Close to the walls there is an out of phase oscillations between these two components. The total energy density (thick curve) shows quantized distribution of the energy inside the box which is due to the known interference effect. The total energy density in the first excited quasiparticle state is depicted in Fig. 8(b). The normalized energy per unit length of box is shown for different orbital energy in Fig. 8(c). It shows almost linear increase for given quantum number as the box length increases which is due to increase in the electron number density. The diffusion current densities are depicted in Fig. 8(d) showing quantized state in this quantity and vanishing at the wall positions. The maximum diffusion current density appears to be in the middle of the box.   

\begin{figure}[ptb]\label{Figure9}
\includegraphics[scale=0.6]{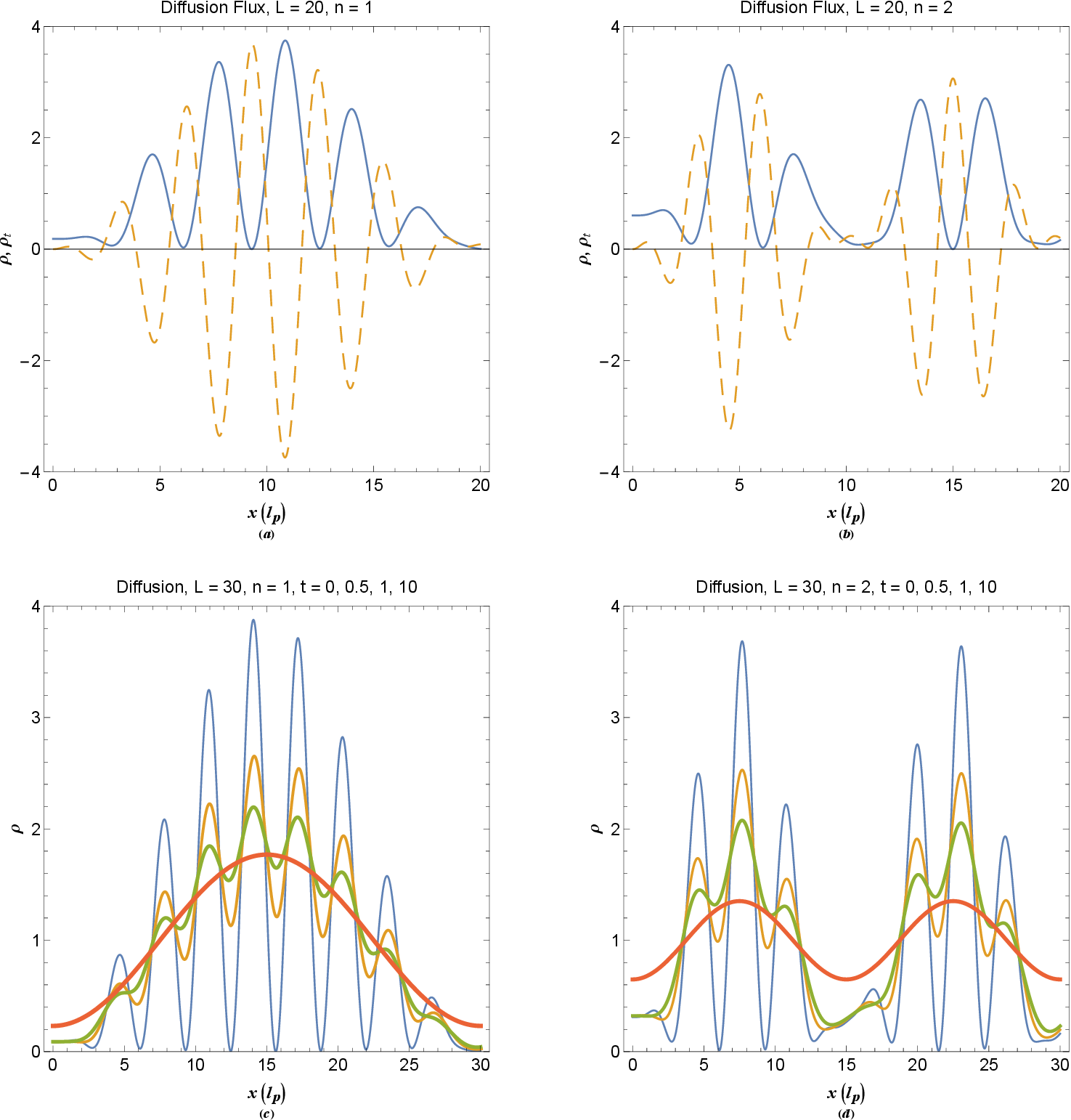}\caption{(a) Generalized density (solid curve) and diffusion density flux (dashed curve) for quasiparticle in a box at ground state level. (b) Generalized density (solid curve) and diffusion density flux (dashed curve) for quasiparticle in a box at first excited state. (c) Variation of generalized density for quasiparticle excitation in a box at ground state level with time. (d) Variation of generalized density for quasiparticle excitation in a box at first excited state with time. The increase in curve thickness indicates the increase in time in plots (c) and (d).}
\end{figure}

Moreover, while the probability current density of stationary states is zero the diffusion current densities do not vanish in the box and are given as
\begin{subequations}\label{dJ1}
\begin{align}
&{J_{\rho \psi }} = \frac{{{A_\rho^2}}}{{4{\alpha ^2}}}\left[ {\cos \left( {{k_1}x} \right) - \cos \left( {{k_2}x} \right)} \right]\left[ {{k_1}\sin \left( {{k_1}x} \right) - {k_2}\sin \left( {{k_2}x} \right)} \right],\\
&{J_{\rho \phi }} = \frac{{{A_\rho^2}}}{{4{\alpha ^2}}}\left[ {k_1^2\cos \left( {{k_2}x} \right) - k_2^2\cos \left( {{k_1}x} \right)} \right]\left[ {{k_2}\sin \left( {{k_1}x} \right) - {k_1}\sin \left( {{k_2}x} \right)} \right],\\
&{J_\rho } = \frac{{{A_\rho^2}}}{{8{\alpha ^2}}}\left[ {{k_1}\left( {1 + k_2^4} \right)\sin \left( {2{k_1}x} \right) + {k_2}\left( {1 + k_1^4} \right)\sin \left( {2{k_2}x} \right)} \right.\\
&\left. { - 4{k_1}\cos \left( {{k_2}x} \right)\sin \left( {{k_1}x} \right) - 4{k_2}\cos \left( {{k_1}x} \right)\sin \left( {{k_2}x} \right)} \right].
\end{align}
\end{subequations} 

\begin{figure}[ptb]\label{Figure10}
\includegraphics[scale=0.6]{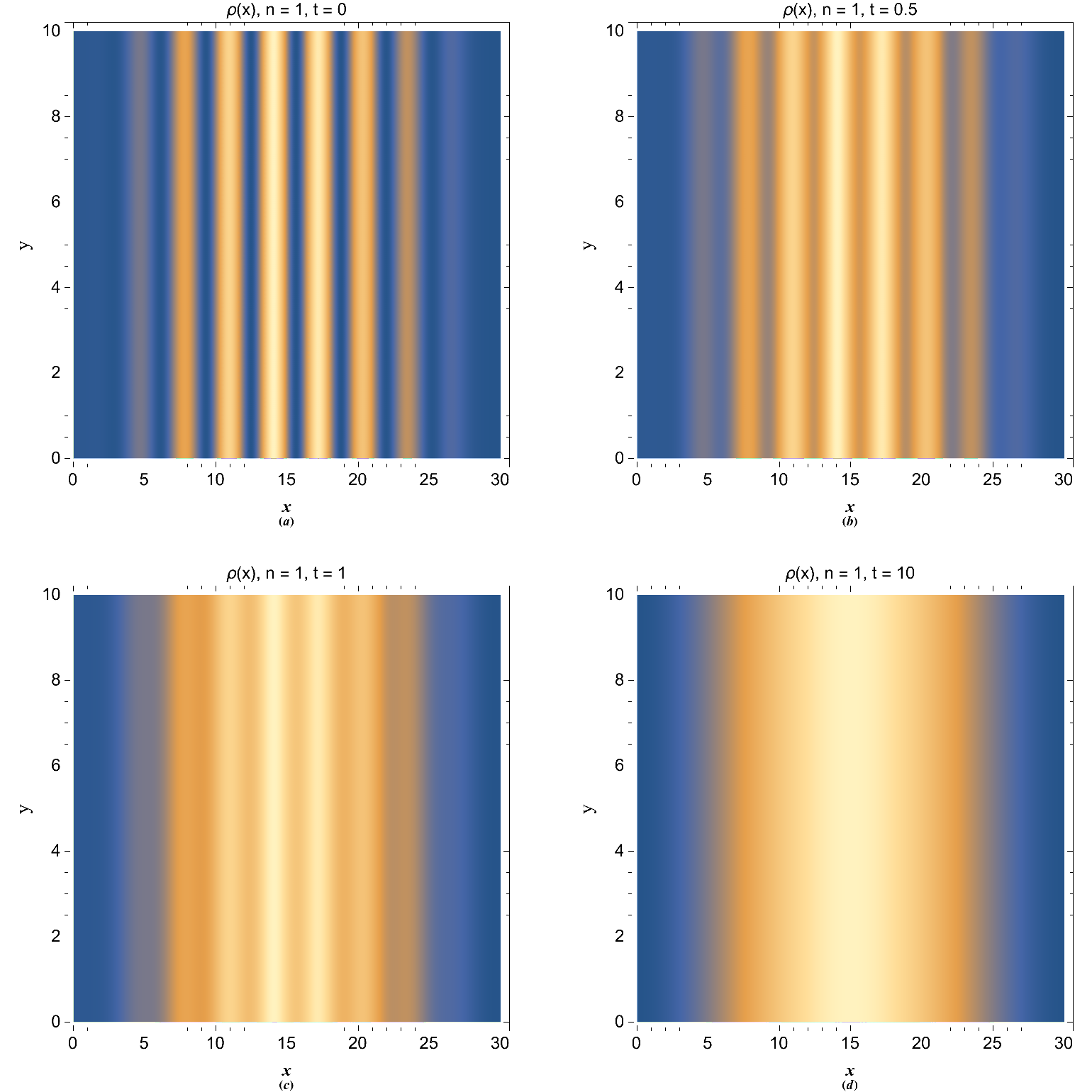}\caption{(a) The density plot of quasiparticle in a box generalized density at ground state level at initial state. (b) Diffusion of the initial state at $t=0.5$. (c) Diffusion of the initial state at $t=1$. (d) Diffusion of the initial state at $t=10$. The time in the unit of inverse of twice the plasmon frequency.}
\end{figure}

The temporal diffusion of the excited states is given by $\partial \rho(x,t)/\partial t=(1/2)\partial^2\rho(x,t)/\partial x^2$ in the absence of the source term. The solution to this equation reads
\begin{subequations}\label{rhot1}
\begin{align}
&\rho (x,t) = \frac{A_\rho^2}{8\alpha^2}\left\{ {2 + k_1^4 + k_2^4 + {{{e}}^{ - 2k_1^2t}}\left( {1 + k_2^4} \right)\cos \left( {2{k_1}x} \right)} \right. + {{{e}}^{ - 2k_2^2t}}\left( {1 + k_1^4} \right)\cos \left( {2{k_2}x} \right)\\
&\left. { - 4{{{e}}^{ - \frac{1}{2}{{\left( {{k_1} - {k_2}} \right)}^2}t}}\cos \left[ {\left( {{k_1} - {k_2}} \right)x} \right] - 4{{{e}}^{ - \frac{1}{2}{{\left( {{k_1} + {k_2}} \right)}^2}t}}\cos \left[ {\left( {{k_1} + {k_2}} \right)x} \right]} \right\},
\end{align}
\end{subequations}
for the initial condition $\rho(x,0)$ given by Eqs. (\ref{norm}). Note that the diffusion in this case is not monotonic but the excited states diffuse to equilibrium value $\rho(x,\infty)=1$ similar to the case of a free quasiparticle excitation. Note also that, first the rapid density oscillations due to single-electron excitations diffuse and then after relatively longer time the equilibrium is reached. 

The diffusion flux along with the corresponding generalized density profile is shown in Figs. 9(a) and 9(b) for given values of box and quantum number. They reveal maximum outward diffusion at density hills and minimum diffusion at density valleys, as expected. The diffusion profiles for different times is shown in Figs. 9(c) and 9(d). It is beautifully shown in Fig. 9(c) that at the ground state quasiparticle level first the excitations due to single electron excitations decay and then the collective density profile diffuses later. The same feature also takes place for the first excited state, as shown in Fig. 9(d). The temporal dynamic of pure state diffusion is shown in a clear density plot view in Fig. 10. 

\section{Energy Density of Damped Quasiparticles}

Finally, we would like to apply the collective quantum diffusion theory to the one-dimensional damped quasiparticle excitations. This case can have quite general application in surface plasmon excitations and the spill-out electron at the half-space electron gas. The non-transient solution in this case with the boundary values $\Psi'_0=\Phi'_0=0$ is \cite{akbdual}
\begin{equation}\label{damped}
\left[ {\begin{array}{*{20}{c}}
{\Phi (x)}\\
{\Psi (x)}
\end{array}} \right] = \frac{{{e^{ - \kappa x}}}}{{2\alpha }}\left[ {\begin{array}{*{20}{c}}
{{\Psi _0} + k_2^2{\Phi _0}}&{ - \left( {{\Psi _0} + k_1^2{\Phi _0}} \right)}\\
{ - \left( {{\Phi _0} + k_1^2{\Psi _0}} \right)}&{{\Phi _0} + k_2^2{\Psi _0}}
\end{array}} \right]\left[ {\begin{array}{*{20}{c}}
\left( {1 + i\kappa /{\beta _1}} \right){{e^{i{\beta _1}x}}}\\
\left( {1 + i\kappa /{\beta _2}} \right){{e^{i{\beta _2}x}}}
\end{array}} \right],
\end{equation}
where $\beta_1=\sqrt{k_1^2-\kappa^2}$ and $\beta_2=\sqrt{k_2^2-\kappa^2}$. We have taken  $\Psi_0=1$ and $\Phi_0=0$ for all simulations.

\begin{figure}[ptb]\label{Figure11}
\includegraphics[scale=0.6]{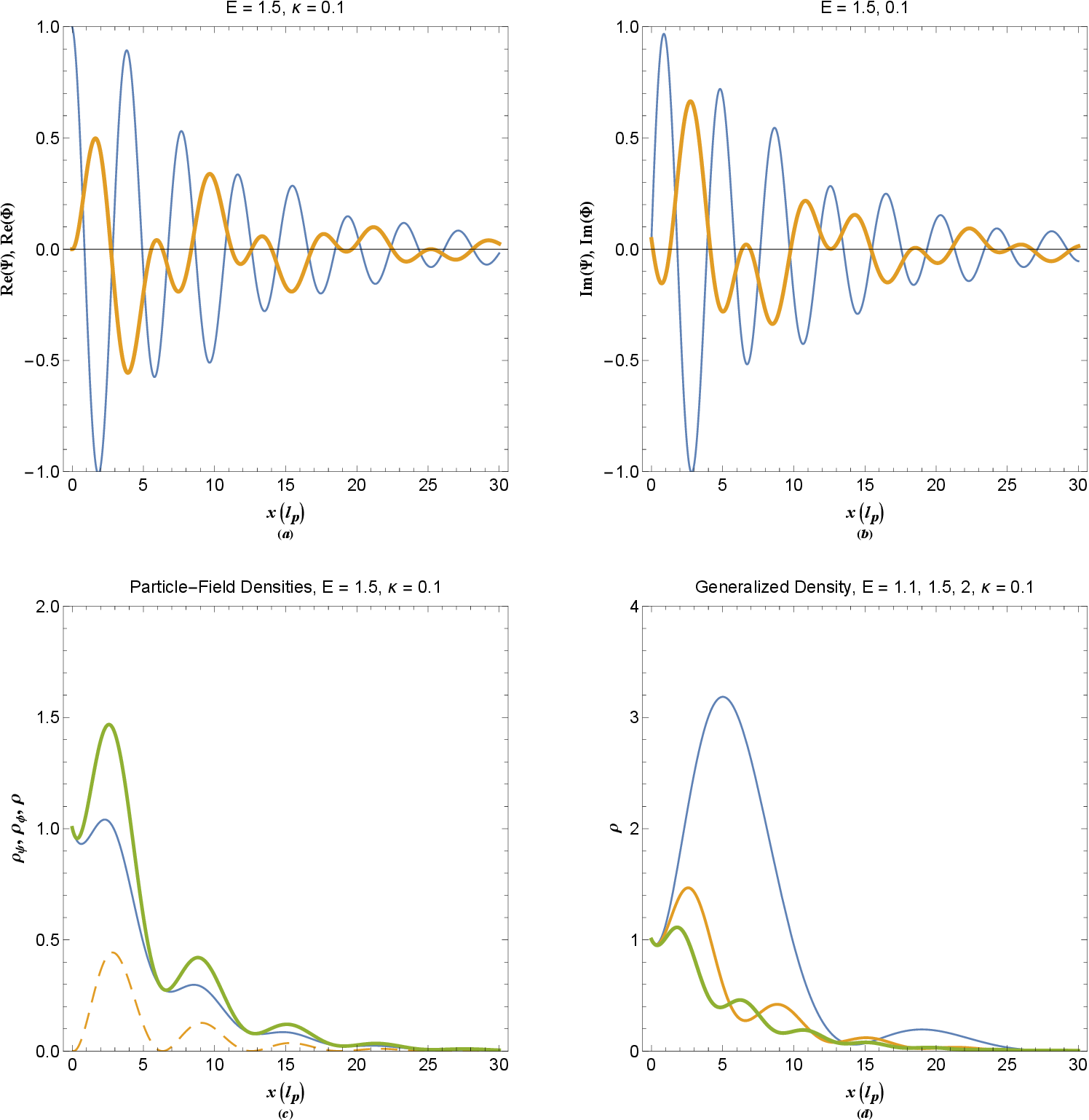}\caption{(a) Real part of damped quasiparticle wavefunction (thin curve) and corresponding electrostatic energy (thick curve) distribution. (b) Imaginary part of damped quasiparticle wavefunction (thin curve) and electrostatic energy (thick curve) profiles. (c) The generalized density due to particle (thin solid curve), field (dashed curve) and total (thick curve). (d) The total generalized density for different quasiparticle energy orbital. Increase in thickness in (d) reflects the increase in varied parameter in this plot. }
\end{figure}

The wavefunction (thin curves) and corresponding electrostatic (thick curves) energy distribution in half space damped quasiparticle excitations are depicted in Figs. 11(a) and 11(b) showing the characteristic decaying dual tone oscillations. The generalized density profiles $\rho_{\psi}$ (solid thin curve) $\rho_\phi$ (dashed) and total $\rho$ (thick curve) are shown in Figs. 11(c). Periodic variations in density profile is apparent due to the interference effect. Such density variations is related to the spill out electron density at plasmonic material surface.     

The generalized density components are given as
\begin{subequations}\label{rho0}
\begin{align}
&{\rho _\psi } = \frac{{{{{e}}^{ - 2\kappa x}}}}{{4{\alpha ^2}\beta _1^2\beta _2^2}}\left\{ {k_1^4\beta _2^2\left( {\beta _1^2 + {\kappa ^2}} \right) + k_2^4\beta _1^2\left( {\beta _2^2 + {\kappa ^2}} \right)} \right.\\
&\left. { + 2{\beta _1}{\beta _2}\left\{ {\kappa \left( {{\beta _2} - {\beta _1}} \right)\sin \left[ {\left( {{\beta _1} - {\beta _2}} \right)x} \right] - \left( {{\beta _1}{\beta _2} + {\kappa ^2}} \right)\cos \left[ {\left( {{\beta _1} - {\beta _2}} \right)x} \right]} \right\}} \right\}\\
&{\rho _\phi } = \frac{{{{{e}}^{ - 2\kappa x}}}}{{4{\alpha ^2}\beta _1^2\beta _2^2}}\left\{ {2\beta _1^2\beta _2^2 + {\kappa ^2}\left( {\beta _1^2 + \beta _2^2} \right)} \right.\\
&\left. { - 2{\beta _1}{\beta _2}\left\{ {\left( {{\beta _1}{\beta _2} + {\kappa ^2}} \right)\cos \left[ {\left( {{\beta _1} - {\beta _2}} \right)x} \right] + \kappa \left( {{\beta _1} - {\beta _2}} \right)\sin \left[ {\left( {{\beta _1} - {\beta _2}} \right)x} \right]} \right\}} \right\}\\
&\rho  = \frac{{{{{e}}^{ - 2\kappa x}}}}{{4{\alpha ^2}\beta _1^2\beta _2^2}}\left\{ {\beta _2^2{\kappa ^2}\left( {k_1^4 + 1} \right) + \beta _1^2\beta _2^2\left( {2 + k_1^4 + k_2^4} \right) + {\kappa ^2}\beta _1^2\beta _2^2\left( {k_2^4 + 1} \right)} \right.\\
&\left. { + 4{\beta _1}{\beta _2}\left\{ {\kappa \left( {{\beta _2} - {\beta _1}} \right)\sin \left[ {\left( {{\beta _1} - {\beta _2}} \right)x} \right] - \left( {{\beta _1}{\beta _2} + {\kappa ^2}} \right)\cos \left[ {\left( {{\beta _1} - {\beta _2}} \right)x} \right]} \right\}} \right\}.
\end{align}
\end{subequations}

\begin{figure}[ptb]\label{Figure12}
\includegraphics[scale=0.6]{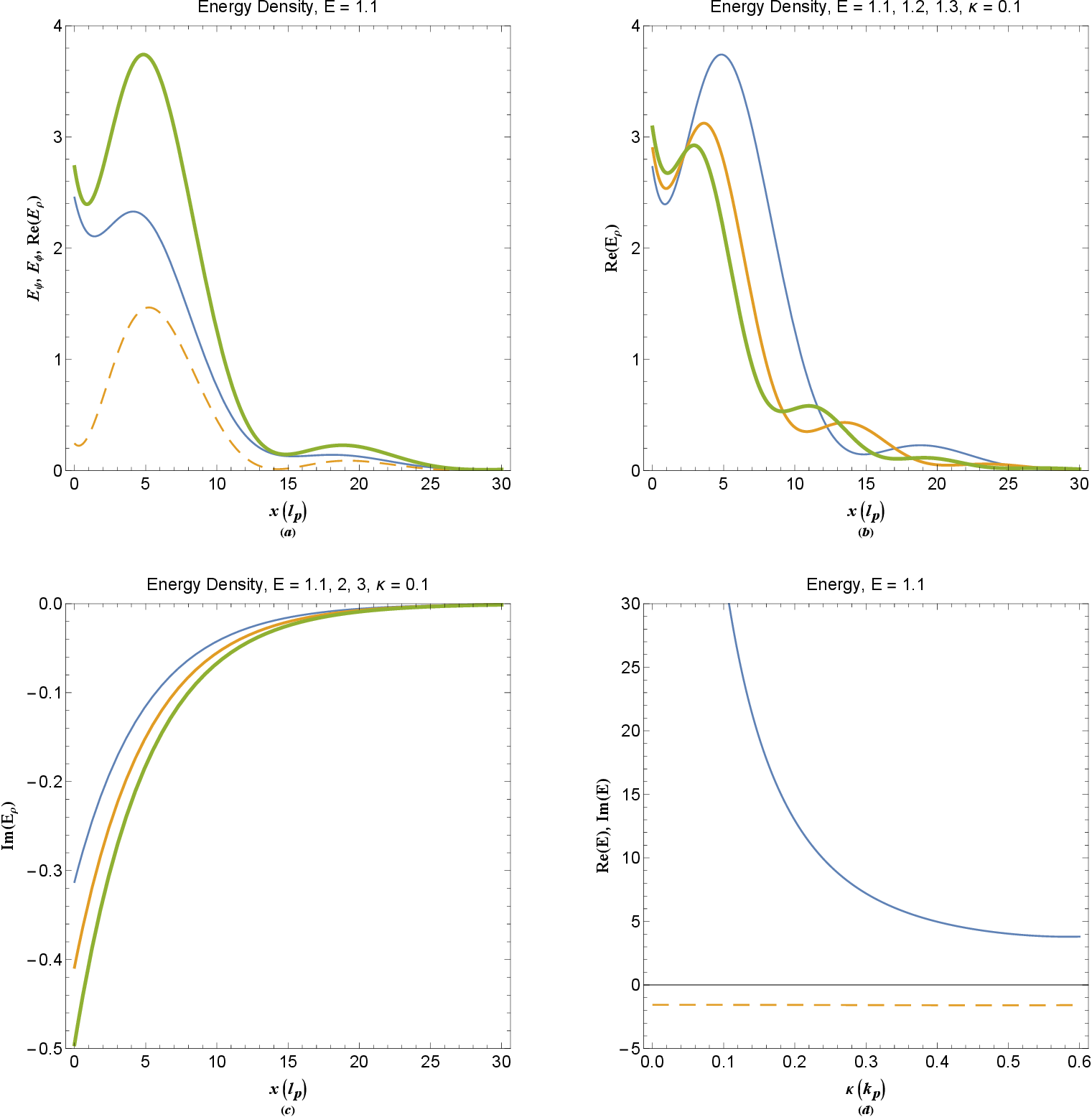}\caption{(a) Energy density of damped quasiparticle excitations due to kinetic (thin solid curve), potential (dashed curve) and total (thick curve). (b) Total real energy density for different orbital energies of damped quasiparticle excitations with curve thickness increasing with increase in the energy. (c) Total imaginary energy density for different orbital energies of damped quasiparticle excitations with curve thickness increasing with increase in the energy. (d) The real and imaginary expected energy of damped quasiparticle excitation with respected to the damping parameter.}
\end{figure}

The energy densities for damped quasiparticle excitation are given as

\begin{subequations}\label{E1}
\begin{align}
&{E_\psi } = \frac{{{{{e}}^{ - 2\kappa x}}}}{{4{\alpha ^2}\beta _1^2\beta _2^2}}\left[ {k_1^2{\beta _2}{{\left( {{\beta _1} - {{i}}\kappa } \right)}^2}{{{e}}^{{{-i}}{\beta _1}x}} - k_2^2{\beta _1}{{\left( {{\beta _2} - {{i}}\kappa } \right)}^2}{{{e}}^{{{-i}}{\beta _2}x}}} \right]\\
& \times \left[ {k_1^2{\beta _2}{{\left( {{\beta _1} + {{i}}\kappa } \right)}^2}{{{e}}^{{{i}}{\beta _1}x}} - k_2^2{\beta _1}{{\left( {{\beta _2} + {{i}}\kappa } \right)}^2}{{{e}}^{{{i}}{\beta _2}x}}} \right],\\
&{E_\phi } = \frac{{{{{e}}^{ - 2\kappa x}}}}{{4{\alpha ^2}\beta _1^2\beta _2^2}}\left[ {{\beta _2}{{\left( {{\beta _1} - {{i}}\kappa } \right)}^2}{{{e}}^{{{ - i}}{\beta _1}x}} - {\beta _1}{{\left( {{\beta _2} - {{i}}\kappa } \right)}^2}{{{e}}^{{{ - i}}{\beta _2}x}}} \right]\\
&\times \left[ {{\beta _2}{{\left( {{\beta _1} + {{i}}\kappa } \right)}^2}{{{e}}^{{{i}}{\beta _1}x}} - {\beta _1}{{\left( {{\beta _2} + {{i}}\kappa } \right)}^2}{{{e}}^{{{i}}{\beta _2}x}}} \right].
\end{align}
\end{subequations} 
The long expression for the total energy density $E_\rho=E_\psi+E_\phi$ is avoided here. Note that the total energy density has imaginary component due to damping. This gives rise to the imaginary energy values which is the characteristic of damping/growing instability of the excitation. The imaginary part of the energy density reads
\begin{equation}\label{iE}
{E_{\rho i}} = \frac{{\kappa {{{e}}^{ - 2\kappa x}}}}{{2{\alpha ^2}{\beta _1}{\beta _2}}}\left[ {{\beta _1}{\beta _2}\left( {{\beta _1} + {\beta _2} - k_1^4{\beta _1} - k_2^4{\beta _2}} \right) + {\kappa ^2}\left( {{\beta _1} + \beta_2  - k_1^4{\beta _2} - k_2^4{\beta _1}} \right)} \right].
\end{equation}
This leads to the total imaginary energy component as
\begin{equation}\label{ie}
{E_i} =  - \frac{{{\beta _1}\left( {k_2^4 - 1} \right)\left( {\beta _2^2 + {\kappa ^2}} \right) + {\beta _2}\left( {k_1^4 - 1} \right)\left( {\beta _1^2 + {\kappa ^2}} \right)}}{{4{\alpha ^2}{\beta _1}{\beta _2}}},
\end{equation}
which indicates a negative constant value characteristics of an exponential decay of the form $\exp(-E_i t)$ with the rate given by (\ref{ie}). The existence of damping in this case originates from the fact that $J_n$ does not vanishes in this case. The components of diffusion current densities are given below
 
Variations in the energy density distribution of damped quasiparticle excitations is shown in Fig. 12. The energy density of damped excitations is imaginary due to the non-vanishing nature of probability current in this case, e.g. see Eq. (\ref{Pd4}). The total energy density (thick curve) shows a pronounced valley close to the boundary ($x=0$). Figure 12(b) shows the total energy density for different orbital energy revealing the fact that with decrease of the energy the energy density valley gets deeper. This is a manifestation of attractive force of the half-space electron gas very close to the surface which is quite similar to the well known Casimir effect. The imaginary part of the energy density is shown to be negative increasing outward the boundary by Fig. 12(c). The real (solid curve) and imaginary (dashed) parts of the total energy $E_t$ at the electron spill out region as varied with the damping parameter is shown in Fig. 12(d). While the real part decreases with the decrease in the damping parameter, which is also related to the collective quantum electron tunneling \cite{akbedge}, the imaginary part is constant negative which indicates a damping effect of the energy-density oscillations with the rate $E_i$.   
 
\begin{figure}[ptb]\label{Figure13}
\includegraphics[scale=0.6]{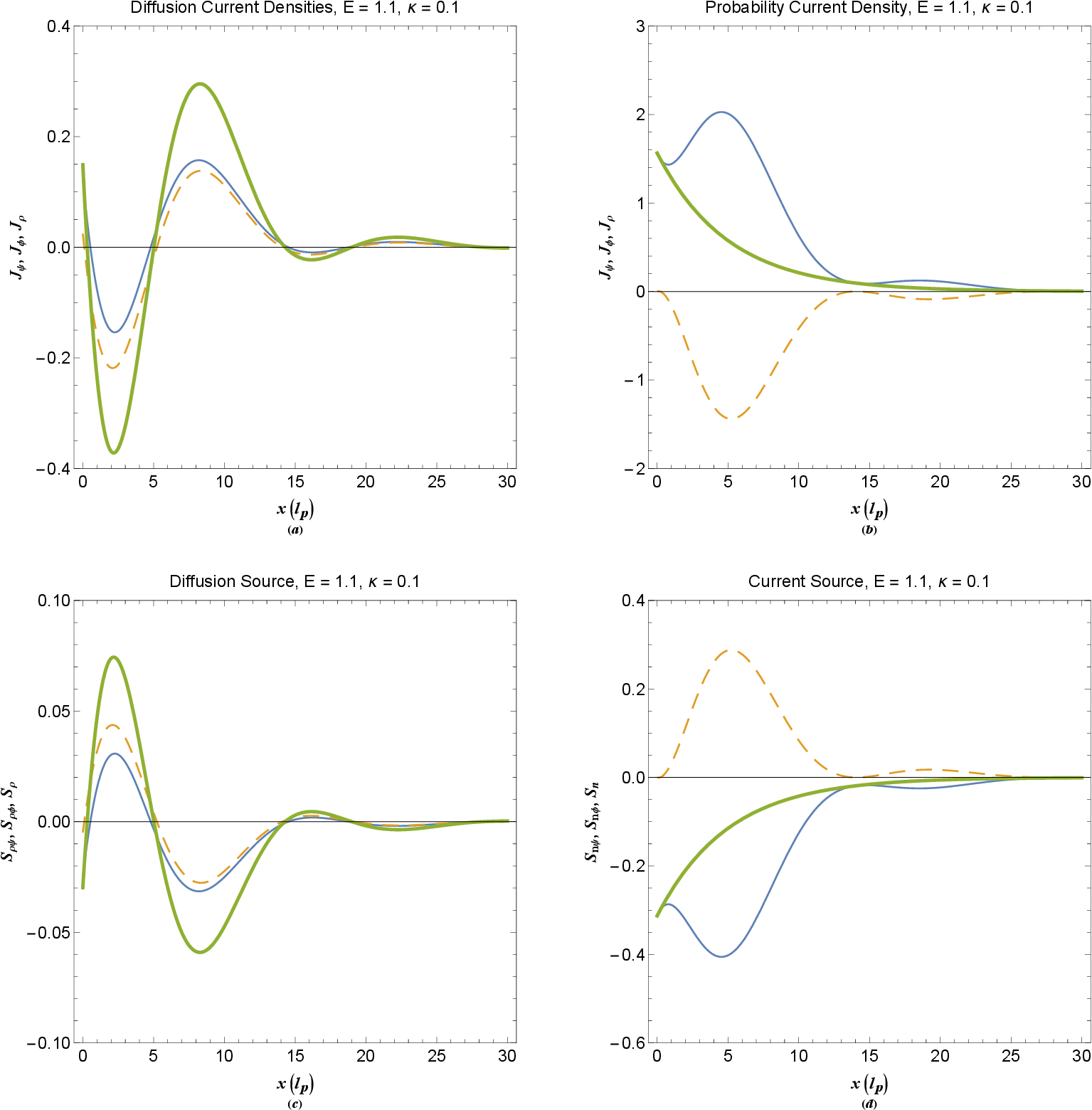}\caption{(a) Variation of diffusion current density of damped quasiparticle due to kinetic (thin solid curve), potential (dashed curve) energies and total diffusion current density (thick curve). (b) Variation of probability current density of damped quasiparticle due to kinetic (thin solid curve), potential (dashed curve) energies and total diffusion current density (thick curve). (c) Variation of diffusion source of damped quasiparticle due to kinetic (thin solid curve), potential (dashed curve) energies and total diffusion current density (thick curve). (d) Variation of probability source of damped quasiparticle due to kinetic (thin solid curve), potential (dashed curve) energies and total diffusion current density (thick curve).}
\end{figure}

\begin{subequations}\label{dJ2}
\begin{align}
&{J_{\rho\psi} } = \frac{{{{{e}}^{ - 2\kappa x}}}}{{4{\alpha ^2}\beta _1^2\beta _2^2}}\left\{ {\kappa k_1^4\beta _2^2\left( {\beta _1^2 + {\kappa ^2}} \right) + \kappa k_2^4\beta _1^2\left( {\beta _2^2 + {\kappa ^2}} \right)} \right.\\
&\left. { + {\beta _1}{\beta _2}\left\{ {\kappa \left( {\beta _1^2 - 4{\beta _1}{\beta _2} + \beta _2^2 - 2{\kappa ^2}} \right)\cos \left[ {\left( {{\beta _1} - {\beta _2}} \right)x} \right] - \left( {{\beta _1} - {\beta _2}} \right)\left( {{\beta _1}{\beta _2} + 3{\kappa ^2}} \right)\sin \left[ {\left( {{\beta _1} - {\beta _2}} \right)x} \right]} \right\}} \right\}\\
&{J_{\rho\phi} } = \frac{{{{{e}}^{ - 2\kappa x}}}}{{4{\alpha ^2}\beta _1^2\beta _2^2}}\left\{ {2\kappa \beta _1^2\beta _2^2 + {\kappa ^3}\left( {\beta _1^2 + \beta _2^2} \right)} \right.\\
&\left. { + \kappa {\beta _1}{\beta _2}\left( {\beta _1^2 - 4{\beta _1}{\beta _2} + \beta _2^2 - 2{\kappa ^2}} \right)\cos \left[ {\left( {{\beta _1} - {\beta _2}} \right)x} \right] - {\beta _1}{\beta _2}\left( {{\beta _1} - {\beta _2}} \right)\left( {{\beta _1}{\beta _2} + 3{\kappa ^2}} \right)\sin \left[ {\left( {{\beta _1} - {\beta _2}} \right)x} \right]} \right\}\\
&{J_\rho } = \frac{{{{{e}}^{ - 2\kappa x}}}}{{4{\alpha ^2}\beta _1^2\beta _2^2}}\left\{ {\beta _2^2{\kappa ^3}\left( {1 + k_1^4} \right) + \kappa \beta _1^2\beta _2^2\left[ {\left( {2 + k_1^4 + k_2^4} \right) + {\kappa ^2}k_2^4} \right]} \right.\\
&\left. {2{\beta _1}{\beta _2}\left\{ {\kappa \left( {\beta _1^2 - 4{\beta _1}{\beta _2} + \beta _2^2 - 2{\kappa ^2}} \right)\cos \left[ {\left( {{\beta _1} - {\beta _2}} \right)x} \right] - \left( {{\beta _1} - {\beta _2}} \right)\left( {{\beta _1}{\beta _2} + 3{\kappa ^2}} \right)\sin \left[ {\left( {{\beta _1} - {\beta _2}} \right)x} \right]} \right\}} \right\}.
\end{align}
\end{subequations} 
The corresponding sources are $S_{\rho\psi}=-2\kappa J_{\rho\psi}$, $S_{\rho\phi}=-2\kappa J_{\rho\phi}$ and $S_{\rho}=-2\kappa J_{\rho}$. The probability current densities are given as follows
\begin{subequations}\label{pJ2}
\begin{align}
&{J_{n\psi} } =\frac{{{{{e}}^{ - 2\kappa x}}\left( {{\beta _1} + {\beta _2}} \right)}}{{4{\alpha ^2}{\beta _1}{\beta _2}}}\left\{ {\left( {{\beta _1}{\beta _2} + {\kappa ^2}} \right)\left\{ {\cos \left[ {x\left( {{\beta _1} - {\beta _2}} \right)} \right] - 1} \right\} + \kappa \left( {{\beta _1} - {\beta _2}} \right)\sin \left[ {\left( {{\beta _1} - {\beta _2}} \right)x} \right]} \right\},\\
&{J_{n\phi} } = \frac{{{{{e}}^{ - 2\kappa x}}}}{{4{\alpha ^2}{\beta _1}{\beta _2}}}\left\{ {{\beta _1}{\beta _2}\left( {{\beta _1} - k_1^4{\beta _1} + {\beta _2} - k_2^4{\beta _2}} \right) + {\kappa ^2}\left( {{\beta _1} - k_2^4{\beta _1} + {\beta _2} - k_1^4{\beta _2}} \right)} \right\},\\
&{J_n } = \frac{{{{{e}}^{ - 2\kappa x}}}}{{4{\alpha ^2}{\beta _1}{\beta _2}}}\left\{ {\left( {{\beta _1} + {\beta _2}} \right)\kappa \sin \left[ {\left( {{\beta _1} - {\beta _2}} \right)x/2} \right]} \right\}\\
&\times \left\{ {\kappa \left( {\beta 2 - \beta 1} \right)\cos \left[ {\frac{{\left( {{\beta _1} - {\beta _2}} \right)}}{2}x} \right] + \left( {{\beta _1}{\beta _2} + {\kappa ^2}} \right)\sin \left[ {\frac{{\left( {{\beta _1} - {\beta _2}} \right)}}{2}x} \right]} \right\}.
\end{align}
\end{subequations}  
The corresponding sources are $S_{n\psi}=-2\kappa J_{n\psi}$, $S_{n\phi}=-2\kappa J_{n\phi}$ and $S_{n}=-2\kappa J_{n}$. The temporal diffusion of the excited states is given by $\partial \rho(x,t)/\partial t=(1/2)\partial^2\rho(x,t)/\partial x^2+\kappa\partial\rho/\partial x$ in the presence of the source term $S_\rho$. The solution to this equation can not be obtained analytically and the numerical analysis of this problem is left for future investigation.

Variations of diffusion and probability current densities due to kinetic (thin solid curve), potential (dashed curve) and total (thick curve) is shown in Figs. 13(a) and 13(b). While the kinetic and potential diffusion current components are closely varying their probability counterparts which appear to contribute to imaginary energy density appear out of phase. The corresponding sources for diffusion and probability current are shown respectivent in Figs. 13(c) and 13(d).  

\section{Concluding Remarks}

In this paper we explored new features of collective quantum excitations in the framework of multistream quasiparticle excitations. We deduced a generalized probability current formula for damped collective quantum excitations and studied the energy density of arbitrary degenerate electron gas. For the first time we proposed the concept collective quantum diffusion in excited quasiparticle states and have shown that all excited levels are unstable and tend to decay into the equilibrium state. The decay rates of the excited states have been calculated for the cases of free quasiparticles and the quasiparticle in a box. This effect is usually attributed to the quantum Landau damping effect. Current study can have important applications in the plasmonics and related fields not only from the technological aspects but also from the theoretical point of view. The model can be generalized to include the external potential and electron spin effect.

\section{Data Availability}

The data that support the findings of this study are available from the corresponding author upon reasonable request.

\end{document}